\input amstex
\documentstyle{amsppt}
\nologo \NoRunningHeads \TagsOnRight \pagewidth{37pc}\pageheight{54pc}
$\qquad\qquad\qquad\qquad\qquad\qquad\qquad\qquad\qquad\qquad\qquad\qquad\qquad\qquad\qquad\qquad\qquad\text{UWThPh-2007-7}$
\bigskip\bigskip\bigskip
\topmatter
\title
Abelian gauge theories on compact manifolds and the Gribov ambiguity
\endtitle
\author
Gerald Kelnhofer
\endauthor
\affil
Faculty of Physics\\University of Vienna\\
Boltzmanngasse 5, A-1090 Vienna\\ Austria
\endaffil
\abstract We study the quantization of abelian gauge theories of principal torus bundles over compact manifolds with and without boundary. It is
shown that these gauge theories suffer from a Gribov ambiguity originating in the non-triviality of the bundle of connections whose geometrical
structure will be analyzed in detail. Motivated by the stochastic quantization approach we propose a modified functional integral measure on the
space of connections that takes the Gribov problem into account. This functional integral measure is used to calculate the partition function,
the Green\rq s functions and the field strength correlating functions in any dimension using the fact that the space of inequivalent connections
itself admits the structure of a bundle over a finite dimensional torus. The Green\rq s functions are shown to be affected by the non-trivial
topology, giving rise to non-vanishing vacuum expectation values for the gauge fields.
\endabstract
\endtopmatter
\document

{\bf 1. Introduction}
\bigskip
Functional integral techniques together with the Faddeev-Popov method [1] play a central role in the quantization of Yang-Mills theories.
Impressive successes of this method were obtained within perturbation theory. The fundamental object in the quantized pure (non-abelian)
Yang-Mills theory is the non-normalized density

$$\Xi ^{(P)} (A)=vol_{\Cal A^{(P)}}\ e^{-S_{inv}(A)},\tag1.1$$
where $vol_{\Cal A^{(P)}}$ denotes the (formal) volume form on the space of all connections $\Cal A^{(P)}$ of a certain principal $G$-bundle $P$
over $M$. $S_{inv}(A)=\frac{1}{2}\int _M tr(F_A\wedge\star F_A)$ is the gauge invariant classical Yang-Mills action, defined by the field
strength $F_A$ and the Hodge star operator $\star$ with respect to a fixed metric on $M$. The trace $tr$ is taken along the Lie-algebra of the
corresponding compact symmetry group $G$. The vacuum expectation value (VEV) of a gauge invariant observable $f\in C^{\infty}(\Cal A^{(P)})$ may
be defined by
$$<f>_P=\frac{\int _{\Cal A^{(P)}}\ \Xi ^{(P)}(A)\cdot f(A)}{\int _{\Cal A^{(P)}}\Xi ^{(P)}(A)}.\tag1.2$$ A difficulty arises because
the integrands appearing in (1.2) are constant along the orbits of the corresponding gauge group $\Cal G^{(P)}$, which have infinite measure.
This implies that the non-physical degrees of freedom must be eliminated before the theory can be quantized. According to the Faddeev-Popov
approach a unique representative is selected from each $\Cal G^{(P)}$-orbit in a smooth way giving rise to a gauge fixing submanifold. The
functional integral over the total space $\Cal A^{(P)}$ is then restricted to this submanifold by extracting the infinite volume of the gauge
group, which is absorbed into an overall normalization constant in the end. As a result the quadratic part of the classical action $S_{inv}$
becomes invertible and the resulting functional integral leads to a consistent perturbative expansion with corresponding Feynman diagrams.
However, the treatment of the infinite gauge group volume is not satisfying from a mathematical point of view.\par

Moreover, in the non-perturbative regime it was envisaged very soon that the Faddeev-Popov formulation suffers from so called Gribov ambiguities
[2]. Topologically this is related to the fact that $\Cal A^{(P)}$ is a non-trivial $\Cal G^{(P)}$-bundle over the gauge orbit space $\Cal
M^{(P)}=\Cal A^{(P)}/\Cal G^{(P)}$ preventing the definition of a smooth global gauge fixing submanifold [3-6]. Gauge fixing is thus only
locally possible and the aim is to find a constructive way to take all relevant gauge fields into account. Not only the non-abelian Yang-Mills
theory suffers from this obstruction but even pure Maxwell theory on the four-dimensional torus $\Bbb T^4$ was shown to be affected by the
Gribov problem using general topological arguments [7].\par

The question whether or how the original Faddeev-Popov approach can be modified appropriately has generated a controversial discussion during
the last years. Several proposals to overcome the Gribov problem have been published (see [8] for a recent overview). In the following outline
we will focus on formulations for an appropriate functional integral in the continuum theory.
\par
One direction follows the original suggestion of Gribov by restricting the functional integral to a submanifold of $\Cal A^{(P)}$, where the
Gribov problem is absent and all gauge fields are uniquely determined [9,10]. Hence the challenge is to find a systematic way to restrict the
Yang-Mills density to this domain of definition and to perform the integration explicitly, see also [11].
\par
A different formulation avoiding the Gribov problem in the Yang-Mills theory has been proposed by some authors in [12,13], where the original
functional integral is modified by the introduction of a non-local gauge fixing term by hand. The modified functional integral restricts the
domain of integration appropriately, yet in this approach the infinite volume of the gauge group has to be omitted.
\par
An alternative way towards the quantization of Yang-Mills theory is to construct a functional integral directly on the gauge orbit space $\Cal
M^{(P)}$ instead of $\Cal A^{(P)}$. The so called "invariant integration" [14,15] relies methodically on the reduction of an integral of
invariant functions over the total space of a finite dimensional principal fibre bundle with compact structure group to an integral over the
base manifold of this bundle multiplied with the finite volume of the symmetry group. If applied literally to the Yang-Mills theory, one would
encounter once again the problem with the ill-defined volume of the gauge group. Hence the idea is to define the partition function of the gauge
theory completely in terms of the resulting functional integral over $\Cal M^{(P)}$. However, compared to the affine space $\Cal A^{(P)}$ the
structure of $\Cal M^{(P)}$ is much more complicated so that an explicit computation of this integral over the gauge orbit space is often not
possible.\par

In [16] the functional integral has been constructed directly on the gauge orbit space $\Cal M^{(P)}$. The basic ingredient is a regularized
Brownian motion governed by the Riemannian structure of $\Cal M^{(P)}$, which is induced by the kinetic term of the (regularized) Yang-Mills
action.\par

A further attempt to be mentioned is [17], where a patching construction for the locally restricted Faddeev-Popov densities - disregarding the
infinite volume of the gauge group - has been investigated.
\par
In this paper we want to present a functional integral measure on the space of all connections that resolves the Gribov problem and provides for
a mathematically reasonable treatment of the gauge degrees of freedom. A novel way to treat these two problems has been proposed some time ago
within the stochastic quantization scheme [18,19]. Generally, the stochastic quantization method of Parisi and Wu [20] was introduced as a new
method for quantizing field theories. It is based on concepts of non-equilibrium statistical mechanics and provides novel and alternative
insights into quantum field theory (see ref. [21], for a comprehensive review and referencing). Let us comment on our proposal in brief: The
gauge fields are regarded as stochastic processes with respect to a fictive so-called "stochastic time", which are governed by an equivalence
class of stochastic differential equation. The notion of equivalence refers to the fact that stochastic correlation functions of gauge invariant
observables are well-defined and unique. This equivalence allows to select a distinguished representative [18,19]. The stochastic scheme can
thus be recast into a formulation in terms of a {\it normalizable\/} probability density as functional of the gauge fields, which has to satisfy
the Fokker-Planck equation [21]. In this respect, the introduction of a damping force along the gauge degrees of freedom regularizing the volume
of the gauge group is one of the main aspects of this approach. The strategy in taking care of the Gribov copies is to restrict the domain of
the stochastic processes to local coordinate patches in the configuration space $\Cal A^{(P)}$, furthermore to construct locally defined
equilibrium distributions and finally to paste them together in such a way that the physical relevant objects become independent of the
particular way this pasting is provided. Distinguished by its concept the whole field content, even the gauge degrees of freedom, has to be
taken into account within the stochastic quantization scheme to permit a probability interpretation.\par

Our aim in this paper is to apply the modified functional integral [19] to abelian gauge theories of connections of principal torus bundles over
$n$-dimensional compact manifolds. These theories are proved to suffer from Gribov ambiguities. So we generalize the results of [7] to a wider
class of manifolds. The motivation to study this theory is twofold: First of all we want to get a more profound understanding of our new concept
for the functional integral by analyzing a simple but non-trivial field theoretical model. As a consequence the interrelation between the
occurrence of the Gribov problem and the necessity for a regularizing measure for the gauge group can be elucidated. However, besides serving as
a laboratory for the new concept the second reason is that abelian gauge theories gained a strong interest during the last years by its own.
Examples are the analysis of two dimensional gauge theories, the description of the fractional quantum Hall effect (see [22] for a comprehensive
review) and questions related to the duality in field theory on three and four dimensional manifolds with and without boundary [23-26]. \par

The paper is structured as follows: In section 2 the concept of the modified functional integral measure will be briefly reviewed. The abelian
field theoretical model which we are going to consider is introduced in section 3. Sections 4 and 5 are devoted to the analysis of the abelian
gauge theory on closed compact manifolds respectively on compact manifolds with a boundary. Since the calculation of the modified functional
integral relies on the knowledge of the bundle geometry of the space of gauge fields, we will analyze its structure in detail in subsection 4.1
for closed manifolds and in subsection 5.1 for manifolds with a non-empty boundary. In both cases, the bundle of connections exhibits a
non-trivial structure which implies that it is impossible to fix the gauge globally. It should be remarked that on closed manifolds the topology
of the gauge orbit space of $\Bbb T^1$-connections has been studied for several years, often in low dimensions. In this respect some of our
results regarding the structure of the gauge group and the topology of the gauge orbit space have already been displayed using different methods
(see e.g. [27]). However, to our knowledge an explicit construction of the various bundle structures in terms of local sections has not appeared
in the literature so far. Our geometrical results will then be used to compute the partition function, the vacuum expectation value of gauge
invariant functions and the Green\rq s functions of the gauge fields for closed manifolds in the subsections 4.2 and 4.3. Analogous results will
be displayed in subsection 5.2 for manifolds with a non-empty boundary. Furthermore our results are compared with those obtained by the
conventional covariant quantization schemes. In section 6 the concept of the modified functional integral and its consequences are illustrated
with two examples, namely the abelian gauge theory on the circle and the abelian gauge theory on two-dimensional closed manifolds. The paper
concludes with a summary of the main results in section 7.
\bigskip\bigskip
{\bf 2. A modified functional integral measure for gauge theories}
\bigskip
In the present publication we want to shed some new light onto the question of how a reasonable partition function can be formulated for a gauge
theory suffering from Gribov ambiguities. In order to ascribe a probabilistic interpretation to the Yang-Mills density according to the
stochastic quantization scheme, the formal measure in (1.1) has to be modified appropriately so that it becomes integrable. The strategy is to
introduce a damping force which regularizes the gauge degrees of freedom. This will be provided by a so called gauge fixing function $S_{gf}$ on
the gauge group $\Cal G^{(P)}$, which is assumed to render the volume of the gauge group\pagebreak

$$Vol(\Cal G^{(P)};S_{gf}):=\int _{\Cal G^{(P)}}\ vol_{\Cal G^{(P)}}\ e^{-S_{gf}}\tag2.1$$
finite. Here $vol_{\Cal G^{(P)}}$ is the (formal) left-invariant volume form on $\Cal G^{(P)}$. In the following we shall consider only gauge
fixing functions $S_{gf}$, which satisfy $Vol(\Cal G^{(P)};S_{gf})=1$.\par

The (non-abelian) gauge group $\Cal G^{(P)}$ acts freely on $\Cal A^{(P)}$, denoted by $(A,g)\mapsto A^g$, so that $\Cal A^{(P)}$ admits the
structure of a principal $\Cal G^{(P)}$-bundle over the gauge orbit space $\Cal M^{(P)}:=\Cal A^{(P)}/\Cal G^{(P)}$ with projection $\pi _{\Cal
A^{(P)}}$. (In fact, the gauge group has to be restricted appropriately to give a free action [4]). The theory is said to possess a Gribov
problem if this bundle is non-trivial. There exists a family of local trivializations $(U_a,\varphi _a)$ given by $ U_a\times\Cal G^{(P)}
@>\varphi _a
>>\pi _{\Cal A^{(P)}}^{-1}(U_a)$ where $\{U_a\}$ is a locally finite open cover of $\Cal M^{(P)}$ and the local diffeomorphisms $\varphi _a(\pi _{\Cal
A}(A),g)=\sigma _a(\pi _{\Cal A^{(P)}}(A))^g$ are generated by a family of local sections $U_a @>\sigma _a
>>\pi _{\Cal A}^{-1}(U_a)$. For the inverse we write
$\varphi _a^{-1}(A)=(\pi _{\Cal A^{(P)}}(A),\omega _a(A))$.
\par
We propose that the quantization of the Yang-Mills theory is described by the following local densities on $\Cal A^{(P)}$

$$\Xi _a^{(P)}=vol_{\Cal A^{(P)}}\mid _{\pi _{\Cal A^{(P)}}^{-1}(U_a)}
\ e^{-S_{inv}-\omega _a^{\ast}S_{gf}}.\tag2.2$$ which - if normalized - appear as equilibrium solutions of the Fokker-Planck operator on each
open set $\pi _{\Cal A^{(P)}}^{-1}(U_a)\subseteq\Cal A^{(P)}$. Due to the Gribov ambiguity these local partition functions must be pasted
together using a partition of unity on the gauge orbit space.

\proclaim{Definition 2.1} Let $\{ p_a \}$ denote a partition of unity on $\Cal M^{(P)}$ subordinate to the open cover $\{U_a\}$. We define a
global (non-perturbative) Yang-Mills density $\Xi ^{(P)}$ by

$$\Xi ^{(P)}:=\sum _{a}(\pi _{\Cal A^{(P)}}^{\ast}p_a)\cdot\Xi _a^{(P)}.\tag2.3$$ Accordingly, the vacuum expectation value (VEV) of a gauge invariant
function $f$ is given by

$$<f>_P=\frac{I^{(P)}(f)}{I^{(P)}(1)},\qquad I^{(P)}(f)=\int _{\Cal A^{(P)}}\Xi ^{(P)}\cdot
f.\tag2.4$$ For the partition function we take $Z^{(P)}:=I^{(P)}(1)$.\endproclaim

It has been shown in [19] that based on this constructive procedure the VEV of gauge invariant observables \roster\item coincides with the
Faddeev-Popov result in the perturbative regime\item is independent of the particular form of the damping force $S_{gf}$ along the gauge group
\item is independent of the particular local trivialization \item is independent of the special choice for the partition of unity.\endroster
This idea to patch the local Yang-Mills densities together to obtain a global functional integral in the field space takes up a suggestion
raised by Singer [3] in his seminal paper.\par

For some applications it is necessary to consider the total configuration space, which consists of disconnected components $\Cal A^{(P)}$
labelled by the equivalence class of bundles $P$. The set of all $\Bbb T^N$ connections over $M$, denoted by $\Cal A^{(M)}$, is given as
disjoint union

$$\Cal A^{(M)}=\bigsqcup _{(P)}\Cal A^{(P)}.\tag2.5$$
Correspondingly, the partition function and the VEV of gauge invariant observables are represented by a sum over equivalence classes of
principal bundles $P$, namely

$$Z=\sum\limits _{(P)}Z^{(P)},\qquad <f>=\frac{\sum\limits_{P}
I^{(P)}(f)}{\sum\limits _{P} I^{(P)}(1)}.\tag2.6$$
\bigskip\bigskip

{\bf 3. The geometrical setting for the abelian gauge theory}
\bigskip
In this section the abelian field theoretical model which we are going to consider in this paper is introduced. As we focus on compact abelian
structure groups only, we can restrict ourselves to the $N$-dimensional torus $\Bbb T^N$ as relevant symmetry group. We shall begin with a brief
review of torus bundles:\par

Let $M$ be a $n$-dimensional connected, oriented and compact manifold with a fixed Riemannian metric. Let us now consider an arbitrary principal
$\Bbb T^N$-bundle $P(M,\pi _P, \Bbb T^N)$ over $M$ with projection $\pi _P$. The group structure on $\Bbb T^N$ is provided by point-wise
multiplication and its Lie algebra $\frak t^N$ is given by $\frak t^N=\sqrt{-1}\ \Bbb R^N$. A $L^2$ inner product can be defined on the complex
$\Omega^k(M;\frak t^N)$ of $k$-forms on $M$ by

$$<\upsilon _1,\upsilon _2>=\sum _{\alpha =1}^N\int _M
\ \upsilon _1^{\alpha}\wedge\star\bar\upsilon _2^{\alpha},\tag3.1$$ where $\star$ is the Hodge star operator with respect to the given metric on
$M$, satisfying $\star ^2=(-1)^{k(n-k)}$ and $\bar\upsilon ^{\alpha}$ denotes the complex conjugate of $\upsilon =(\upsilon ^{\alpha})_{\alpha
=1}^N\in\Omega^k(M;\frak t^N)$.\par

The $C^{\infty}$-Hilbert manifold of all connections on $P$ of a certain Sobolev class will be denoted by $\Cal A^{(P)}$. The gauge group $\Cal
G^{(M)}$ is defined as the group of vertical bundle automorphisms on $P$ and can be identified with the Hilbert Lie-Group $C^{\infty}(M,\Bbb
T^N)$ of differentiable maps between $M$ and $\Bbb T^N$. Finally, its Lie-algebra $\frak G^{(M)}$ is given by $\frak G^{(M)}=C^{\infty}(M;\frak
t^N)$.\par

Under an arbitrary gauge transformation $g\in\Cal G^{(M)}$, the gauge fields transform according to

$$\Cal A\mapsto A^g=A+(\pi _P^{\ast}g)^{\ast}\vartheta
\qquad g\in\Cal G^{(M)},\tag3.2$$ where $\vartheta\in\Omega ^1(\Bbb T^N;\frak t^N)$ is the Maurer Cartan form on $\Bbb T^N$. (For notational
convenience we shall not distinguish between $\pi _P^{\ast}g$ and $g$.) \par

How can torus bundles be classified? The topological type of $\Bbb T^N$ torus bundles is expressed by the first Cech-cohomology $H^1(M;sh_M(\Bbb
T^N))$, where $sh_M(\Bbb T^N)$ denotes the sheaf of all $\Bbb T^N$ valued differentiable functions on $M$. The sheaves of $\Bbb Z^N$ and $\Bbb
R^N$ valued differentiable functions on $M$, which are denoted by $sh_M(\Bbb Z^N)$, $sh_M(\Bbb R^N)$, respectively, fit into the following exact
sequence of sheaves

$$0\rightarrow sh_M(\Bbb Z^N)\rightarrow
sh_M(\Bbb R^N)\rightarrow sh_M(\Bbb T^N)\rightarrow 1,\tag3.3$$

which induces a corresponding long exact sequence in cohomology

$$\dots\rightarrow \hat H^1(M, sh_M(\Bbb R^N))\rightarrow\hat
H^1(M, sh_M(\Bbb T^N))\rightarrow\hat H^2(M, sh_M(\Bbb Z^N))\rightarrow\hat H^2(M, sh_M(\Bbb R^N))\rightarrow\ldots\ \tag3.4$$ Since the sheaf
$sh_M(\Bbb R^N)$ is fine, the set $\frak P[M,\Bbb T^N]$ of equivalence classes of principal $\Bbb T^N$ bundles over $M$ is given by

$$\frak P[M;\Bbb T^N]=\hat H^1(M, sh_M(\Bbb T^N))=H^2(M,\Bbb Z^N)=
\bigoplus\limits _{i=1}^N H^2(M,\Bbb Z),\tag3.5$$ so that any principal $\Bbb T^N$-bundle is classified by an integer cohomology class $c\in
H^2(M,\Bbb Z^N)$. Accordingly, $c=c_1\oplus\cdots\oplus c_N$, where each component $c_{\alpha}\in H^2(M,\Bbb Z)$ determines a principal circle
bundle $P^{\alpha}(M,\Bbb T^1)$ over $M$ having $c_{\alpha}$ as its first Chern class. Thus $P$ can be equivalently viewed as $N$-fold fiber
product $P^1\times _M\times\cdots\times _MP^N$ over $M$.
\par
Let $F_A=(F_A^{\alpha})_{\alpha =1}^{N}\in\Omega ^2(M;\frak t^N)$ denote the field strength of the $\Bbb T^N$-connection $A$ on $P$. Each
component $F_A^{\alpha}$ can be regarded as field strength of the $\alpha$-th principal $\Bbb T^1$-bundle $P^{\alpha}$ in the fiber product $P$.
The classical gauge invariant action is defined by

$$S_{inv}(A)=\frac{1}{2}\sum _{\alpha ,\beta =1}^N\int _M\lambda
_{\alpha\beta}F_A^{\alpha}\wedge\star\bar F_A^{\beta},\tag3.6$$ where $(\lambda _{\alpha\beta})_{\alpha ,\beta =1}^N$ is a symmetric positive
definite matrix with $\det\lambda =1$. This matrix determines the relative couplings between the components $A^{\alpha}$ of the $\Bbb T^N$-gauge
fields $A$ on $P$. From a physical point of view some extensions of (3.6) are of particular interest: If the conventional Maxwell action is
extended by a theta term the resulting partition function was shown to exhibit a non-trivial transformation behavior under electric-magnetic
duality [23-26]. On the other hand, if the action (3.6) is extended by an additional Chern-Simons term in a three dimensional space-time, this
model allows for a mathematical description of the fractional quantum Hall effect. The integer resulting from the evaluation of the
corresponding Chern classes $c^{\alpha}$ ($\alpha =1,\ldots N$) along the 2-dimensional space admits the interpretation of the total number of
electrons in the $\alpha$-th Landau level [22].\par

Provided by the matrix of couplings there is a second $L^2$ inner product on the complex $\Omega (M;\frak t^N)$ given by

$$<\upsilon _1,\upsilon _2>_{\lambda}=\sum _{\alpha ,\beta =1}^N
\int _M\lambda _{\alpha\beta}\upsilon _1^{\alpha}\wedge\star\bar\upsilon _2^{\beta},\tag3.7$$ where $\upsilon =(\upsilon ^1,\ldots ,\upsilon
^N)\in\Omega ^k(M;\frak t^N)$.
\bigskip\bigskip

{\bf 4. Abelian gauge theories on closed manifolds}\bigskip

In this chapter we want to construct the modified functional integral for the abelian gauge theory on closed manifolds. We begin with an
analysis of the geometrical properties of the gauge group. Based on these considerations we will then derive two results regarding the bundle
structure of the space of connections.
\bigskip
{\bf 4.1. The geometry of the abelian gauge fields}
\bigskip
The action (3.2) of the gauge group $\Cal G^{(M)}$ is not free possessing the non-trivial isotropy group $\Bbb T^N$, namely the subgroup of
constant gauge transformations. In order to get a free action let us now choose an arbitrary but fixed reference point $x_0\in M$. By
restricting the gauge group to the subgroup $\Cal G _{\ast}^{(M)}=\{g\in\Cal G|g(x_0)=1\}$ which itself is diffeomorphic to $\Cal G^{(M)} /\Bbb
T^N$ by $g\rightarrow g\cdot g(x_0)$, we finally obtain a free action of $\Cal G _{\ast}^{(M)}$ on $\Cal A^{(P)}$. This gives rise to a smooth
gauge orbit space $\Cal M _{\ast}^{(P)}=\Cal A^{(P)} /\Cal G _{\ast}^{(M)}$, which has to be regarded as the true configuration space of the
theory. For the Maxwell theory ($N=1$) some of the results regarding the gauge group topology have been considered in [28].\par

Let us denote by $Z_k(M;\Bbb Z)$ the subcomplex of all closed smooth singular $k$-cycles on $M$. We define the abelian group
$$\Omega _{\Bbb Z}^k(M,\Bbb R^N)=\{\alpha\in\Omega ^k(M;\Bbb R^N)
\vert\quad d\alpha =0,\quad \int _{\gamma}\alpha \in\Bbb Z^N\quad \forall\gamma\in Z_k(M;\Bbb Z)\}\tag4.1.1$$ of all closed $\Bbb R^N$-valued
differential $k$-forms with integer periods and denote by $H_{\Bbb Z}^k(M;\Bbb R^N)$ the corresponding cohomology group.
\par
The question of how the subgroup of constant gauge transformations is related to the gauge group is answered by the following statement:

\proclaim{Proposition 4.1} The following sequence of abelian groups is split exact
$$0\rightarrow\Bbb T^N \rightarrow\Cal G^{(M)} @>\kappa _{(M)} >>\Omega _{\Bbb Z}^1(M,\Bbb
R^N)\rightarrow 0\qquad \kappa _{(M)}(g)=\frac{1}{2\pi\sqrt{-1}}g^{\ast}\vartheta .\tag4.1.2$$
\endproclaim\pagebreak

\demo{Proof} The split is given by the isomorphism of abelian groups

$$\split &\tilde{\kappa} _{(M)}\colon \Omega _{\Bbb Z}^1(M,\Bbb R^N)\times\Bbb T^N\rightarrow\Cal
G^{(M)}\\
&\tilde{\kappa}_{(M)}(\alpha ,t)(x)=t\cdot\exp{2\pi\sqrt{-1}\int _{c_x}\alpha
}=t\cdot\exp{2\pi\sqrt{-1}\int _0^1\ c_x^{\ast}\alpha}\\
& \tilde{\kappa} _{(M)}^{-1}(g)=(\kappa _{(M)}(g),g(x_0)),\endsplit\tag4.1.3$$

where $c_x\colon [0,1]\rightarrow M$ is a path in $M$ connecting $x_0$ with $x$. That this integral is already well-defined can be seen by
choosing a different path $c_x^{\prime}$ connecting $x_0$ and $x$. Since the combined path $c_x^{\prime}\diamond c_x$ can be regarded as element
in $Z_1(M;\Bbb Z)$. The integration of any element in $\Omega _{\Bbb Z}^1(M,\Bbb R^N)$ along this cycle gives an integer. \qed\enddemo

The co-differential $d_k^{\ast}=(-1)^{n(k+1)+1}\star d_{n-k}\star\colon\Omega ^{k}(M;\Bbb R)\rightarrow\Omega ^{k-1}(M;\Bbb R)$ gives rise to
the Laplacian operator $\Delta _k =d_{k+1}^{\ast}d_k+d_{k-1}d_{k}^{\ast}$. Let $Harm^{k}(M)^{\bot}$ denote the orthogonal complement of the
space of harmonic $k$-forms $Harm^k(M)$ with values in $\Bbb R$, then we can define the Green´s operator [29]

$$G_k\colon\Omega ^k(M;\Bbb R)\rightarrow Harm^{k}(M)^{\bot},\quad G_k=
(\Delta _k\vert _{Harm^{k}(M)^{\bot}})^{-1}\circ \Pi ^{Harm^{k}(M)^{\bot}},\tag4.1.4$$ where $\Pi ^{Harm^{k}(M)^{\bot}}$ is the projection of
$\Omega ^k(M;\Bbb R)$ onto $Harm^{k}(M)^{\bot}$. By construction $\Delta _k\circ G_k=G_k\circ\Delta _k =\Pi ^{Harm^{k}(M)^{\bot}}$.\par

It is evident that the Lie algebra $\frak G_{\ast}^{(M)}$ of the restricted gauge group $\Cal G_{\ast}^{(M)}$ consists of those $C^{\infty}$
maps from $M$ to $\frak t^N$, which vanishes in $x_0$. The next result shows that the pointed gauge group $\Cal G_{\ast}^{(M)}$ is not
connected.

\proclaim{Proposition 4.2} The following sequence of abelian groups is split exact

$$0\rightarrow\frak G_{\ast}^{(M)}@>\exp >>\Cal G_{\ast}^{(M)}
@>\kappa _{(M)}^{\prime}>> H_{\Bbb Z}^1(M;\Bbb R^N)\rightarrow 0,\tag4.1.5$$ where $\kappa _{(M)}^{\prime}(g)=[\kappa _{(M)}(g)]$.\endproclaim
\demo{Proof} It is easy to see that the exponential function $\exp$ is indeed a monomorphism. A split of (4.1.5) is given by the following
isomorphism of abelian groups

$$\split &\hat\kappa _{(M)}\colon H_{\Bbb Z}^1(M;\Bbb R^N)\times\frak
G_{\ast}^{(M)}\rightarrow\Cal G_{\ast}^{(M)}\\ &\hat\kappa _{(M)}([\alpha ],\xi )(x)=\exp{(2\pi\sqrt{-1}\int _{c_x}\Pi ^{Harm^1(M)}(\alpha
))}\cdot\exp\xi (x)
\\ &\hat\kappa _{(M)}^{-1}(g)=(\kappa
_{(M)}^{\prime}(g), G_0d_1^{\ast}g^{\ast}\vartheta - (G_0d_1^{\ast}g^{\ast}\vartheta )(x_0)).\endsplit\tag4.1.6$$ \qed\enddemo

Now we will prove that even an abelian gauge theory would admit a Gribov ambiguity if the space time manifold $M$ is topologically non-trivial.
This generalizes the previous result [7], where the existence of Gribov ambiguities has been shown for Maxwell theory on the four-torus.

\proclaim{Theorem 4.3} $\Cal A^{(P)}$ is a flat principal bundle over $\Cal M_{\ast}^{(P)}$ with structure group $\Cal G_{\ast}^{(M)}$ and
projection $\pi _{\Cal A^{(P)}}$. This bundle is trivializable if $H^1(M;\Bbb Z)=0$.
\endproclaim

\demo{Proof} We are going to construct a bundle atlas explicitly. For this we have to define an open cover of the gauge orbit space and a family
of local sections. For any fixed $l,N\in\Bbb N$ we consider the exact sequence of abelian groups

$$0\rightarrow\Bbb Z^{lN}\rightarrow\Bbb R^{lN}@>\exp{2\pi\sqrt{-1}(.)}>>\Bbb
T^{lN}\rightarrow 0,\tag4.1.7$$ which gives the universal covering of the $lN$-dimensional torus $\Bbb T^{lN}$. Let us view $\Bbb T^{lN}$ as the
product

$$\Bbb T^{lN}=\undersetbrace l \to{\Bbb T^N\times\cdots\times\Bbb
T^N}=\undersetbrace l\to{ (\undersetbrace N \to{\Bbb T^1\times\cdots\times\Bbb T^1})\times\cdots\times (\undersetbrace N \to{\Bbb
T^1\times\cdots\times\Bbb T^1})}.\tag4.1.8$$ We introduce an open cover $\Cal V$ of $\Bbb T^{lN}$ by the following family of open sets

$$\Cal V=\{V_{a}\vert\quad
a:=(a_1,\ldots ,a_j,\ldots ,a_{l}),\quad a_j:=(a_{j1},\ldots ,a_{j\alpha},\ldots ,a_{jN}), a_{j\alpha}\in\Bbb Z_2=\{1,2\}\},\tag4.1.9$$ where
$V_a=V_{a_1}\times\cdots\times V_{a_j}\times\cdots\times V_{a_{l}}$ is a open set in $\Bbb T^{lN}$. Each $V_{a_j}$ is itself the product of open
sets $V_{a_j}=V_{a_{j1}}\times\cdots\times V_{a_{j\alpha}}\times\cdots\times V_{a_{jN}}$ in the $k$-th $N$-dimensional torus $\Bbb T^N$ within
(4.1.8). Here $V_{1}= \Bbb T^1\backslash\{northern pole\}$ for $a_{j\alpha}=1$ and $V_{2}= \Bbb T^1\backslash\{southern pole\}$ for
$a_{j\alpha}=2$ provide an open cover of each 1-torus $\Bbb T^1$. Let us choose the following two local sections of the universal covering $\Bbb
R^1 \rightarrow \Bbb T^1$

$$\align & s_{a_{j\alpha}}(z)=\{ {\frac{1}{2\pi }\arccos |_{(0,\pi ]}\Re z\atop
\frac{1}{2\pi }\arccos |_{[\pi,2\pi )}\Re z }\qquad {\Im z\geq 0\atop \Im z<0},\quad a_{j\alpha}=1 \\
& s_{a_{j\alpha}}(z)=\{ {\frac{1}{2\pi }\arccos |_{(\pi ,2\pi ]}\Re z\atop \frac{1}{2\pi }\arccos |_{[2\pi,3\pi )}\Re z}\qquad {\Im z\leq 0\atop
\Im z>0}\quad a_{j\alpha}=2,\tag4.1.10\endalign$$ where $z=\Re z+\sqrt{-1}\Im z\in\Bbb T^1$. The corresponding transition functions
$g_{a_{j\alpha}a_{j\alpha}^{\prime}}^{\Bbb T^{1}}\colon V_{a_{j\alpha}}\cap V_{a_{j\alpha}^{\prime}}\rightarrow\Bbb Z$ are given by
$$s_{a_{j\alpha}^{\prime}}(z_{j\alpha})=s_{a_{j\alpha }}(z_{j\alpha})+
g_{a_{j\alpha}a_{j\alpha}^{\prime}}^{\Bbb T^{1}}(z_{j\alpha}).\tag4.1.11$$ Evidently a family of $2^{lN}$ local sections $s_a\colon
V_a\subset\Bbb T^{lN}\rightarrow\Bbb R^{lN}$ can be induced by

$$s_a=(s_{a_1},\cdots ,s_{a_l})=\left( (s_{a_{11}},\cdots ,s_{a_{1N}}),
\cdots ,(s_{a_{l1}},\cdots ,s_{a_{lN}}) \right),\tag4.1.12$$ where on $V_a\cap V_{a^{\prime}}$ the corresponding sections $s_a$ and
$s_{a^{\prime}}$ are related by the locally constant transition functions $g_{aa^{\prime}}^{\Bbb T^{Nl}}\colon V_a\cap
V_{a^{\prime}}\rightarrow\Bbb Z^{Nl}$
$$\multline g_{aa^{\prime}}^{\Bbb T^{Nl}}(\vec{z}_{1},\ldots ,\vec{z}_{l})=
( g_{a_{1}a_{1}^{\prime}}^{\Bbb T^{N}}(\vec{z}_{1}),\ldots , g_{a_{j}a_{j}^{\prime}}^{\Bbb T^{N}}(\vec{z}_{j }),\ldots
,g_{a_{l}a_{l}^{\prime}}^{\Bbb T^{N}}(\vec{z}_{l}))= \\
=((g_{a_{11}a_{11}^{\prime}}^{\Bbb T^{1}}(z_{11}),\ldots ,g_{a_{1N}a_{1N}^{\prime}}^{\Bbb T^{1}}(z_{1N})),\ldots
,(g_{a_{l1}a_{l1}^{\prime}}^{\Bbb T^{1}}(z_{l1}),\ldots ,g_{a_{lN}a_{lN}^{\prime}}^{\Bbb T^{1}}(z_{lN}))),
\endmultline\tag4.1.13$$
for $\vec{z}_j=(z_{j1},\ldots ,z_{jN})\in \Bbb T^N$ with $j=1,\ldots , l$. These local sections will be the building blocks for the construction
of a bundle atlas.\par

Let $Harm_{\Bbb Z}^k(M;\Bbb R)$ denote the abelian group of harmonic $k$-forms with integer periods and let $D_{n-1}\colon H^{n-1}(M;\Bbb
Z)\rightarrow H_{1}(M;\Bbb Z)$, $D_{n-1}(\nu )=\nu\cap [M]$ be the Poincar\'e duality isomorphism [30]. Here $\cap$ is the cap product and $[M]$
denotes the fundamental cycle.
\par
Since the homology of $M$ is finitely generated with rank $b_1$ (the first Betti number of $M$) we shall choose a set of 1-cycles $\gamma _i\in
Z_1(M,\Bbb Z)$, $i=1,\ldots ,b_1$, whose homology classes $[\gamma _i]$ provides a Betti basis thus generating the free part $H_1(M;\Bbb
Z)/TorH_1(M;\Bbb Z)$ in $H_1(M;\Bbb Z)$. Here $TorH_1(M;\Bbb Z)$ denotes the torsion part of the first homology group. Then
$D_{n-1}^{-1}([\gamma _i])$ provides a basis for cohomology, from which a basis of harmonic forms $(\rho _i^{(n-1)})_{i=1}^{b_{n-1}}\in
Harm_{\Bbb Z}^{n-1}(M;\Bbb R)$ can be selected according to the following isomorphisms

$$H^{n-1}(M;\Bbb Z)/TorH^{n-1}(M;\Bbb Z)\cong H_{\Bbb Z}^{n-1}(M;\Bbb R) \cong Harm_{\Bbb
Z}^{n-1}(M;\Bbb R).\tag4.1.14$$ Using the Poincar\'e duality and the Universal Coefficient Theorem it follows that the product

$$\split H^1(M;\Bbb Z)/TorH^1(M;\Bbb Z) &\times H^{n-1}(M;\Bbb Z)/TorH^{n-1}(M;\Bbb Z)
\rightarrow\Bbb Z\\  (\mu ,\nu )&\mapsto <\mu ,D_{n-1}(\nu )>=<\mu\cup\nu,[M]>,\endsplit\tag4.1.15$$ gives a perfect pairing [30], where $<,>$
denotes the evaluation in cohomology. We remark that $TorH^1(M;\Bbb Z)=0$. A basis $(\rho _i^{(1)})_{i=1}^{b_1}\in Harm_{\Bbb Z}^1(M;\Bbb R)$
can be adjusted in such a way so that

$$\int _{\gamma _j}\rho _i^{(1)} =\int _M\ \rho _i^{(1)}\wedge\rho _j^{(n-1)}=\delta
_{ij}.\tag4.1.16$$ Hence $\int _{\gamma _j}\alpha =\int _M\alpha\wedge\rho _j^{(n-1)}$ holds for any $[\alpha ]\in H^1(M;\Bbb R)$. On
$Harm^1(M;\Bbb R)$ there exists an induced metric
$$h_{jk}=<\rho _{j}^{(1)},\rho _{k}^{(1)}>.\tag4.1.17$$
For any choice of an arbitrary but fixed background gauge field $A_0\in\Cal A^{(P)}$ there exists a smooth surjective map $\pi _{\Cal
M_{\ast}^{(P)}}^{A_0}\colon\Cal M_{\ast}^{(P)}\rightarrow\Bbb T^{b_1N}$ defined by

$$\pi _{\Cal M_{\ast}^{(P)}}^{A_0}([A])=(e^{\int _M (A-A_0)\wedge\rho
_1^{(n-1)}},\ldots , e^{\int _M (A-A_0)\wedge\rho _{b_1}^{(n-1)}}),\tag4.1.18$$ where its components can be rewritten in terms of the inner
product (3.1), namely

$$\int _M (A-A_0)\wedge\rho _j^{(n-1)}=(-1)^n<A-A_0,\star\rho _j^{(n-1)}>.\tag4.1.19$$
The family of open sets $U_a^{A_0}=(\pi_{\Cal M_{\ast}^{(P)}}^{A_0})^{-1}(V_a)$ provides a finite open cover $\Cal U^{A_0}=\{U_a^{A_0}\}$ of the
infinite dimensional manifold $\Cal M_{\ast}^{(P)}$. Now we can construct a bundle atlas from the family of local trivializations $\varphi
_{a}^{A_0}:U_{a}^{A_0}\times\Cal G_{\ast}^{(M)}\rightarrow(\pi _{\Cal A^{(P)}})^{-1}(U_a^{A_0})$, $\varphi _{a}^{A_0}([A],g)=A^{(\omega
_{a}^{A_0}(A))^{-1}g}$, and $(\varphi _{a}^{A_0})^{-1}(A)=(\pi _{\Cal A^{(P)}}(A),\omega _{a}^{A_0}(A))$, where

$$\split &\omega _{a}^{A_0}\colon\pi _{\Cal A^{(P)}}^{-1}(U_{a}^{A_0})\rightarrow \Cal
G_{\ast}^{(M)}\\ & \omega _{a}^{A_0}(A) = \hat\kappa _{M}([\sum _{j=1}^{b_1}\epsilon _{a_j}(A)\rho _k^{(1)}],
  \exp{G_0 d_1^{\ast}(A-A_0)}\cdot\exp{G_0 d_1^{\ast}(A-A_0)(x_0)}),\\
&\epsilon _{a_j }^{A_0}=(\epsilon _{a_{j1}}^{A_0},\ldots ,\epsilon _{a_{j\alpha }}^{A_0},\ldots ,\epsilon _{a_{jN}}^{A_0})
\colon\pi _{\Cal A^{(P)}}^{-1}(U_{a})\rightarrow\Bbb Z^{N}\\
&\epsilon _{a_{j\alpha } }^{A_0}(A) =\frac{1}{2\pi\sqrt{-1}}\int _M (A^{\alpha}-A_0^{\alpha})\wedge\rho _j^{(n-1)}- s_{a_{j\alpha}}(\exp{\int _M
(A^{\alpha}-A_0^{\alpha})\wedge\rho _j^{(n-1)}}),\endsplit\tag4.1.20$$ for $\alpha =1,\ldots ,N$. To verify that (4.1.20) indeed gives a local
trivialization of the bundle, we recognize that $\frac{1}{2\pi\sqrt{-1}}\int _{\gamma _j}\ g^{\ast}\vartheta =:m_j\in\Bbb Z^N$. With respect to
the basis $(\rho _{j}^{(1)})_{j=1}^{b_1}$, the orthogonal projector onto $Harm^1(M)$ reads

$$\Pi ^{Harm^1(M)}(\alpha )=\sum _{j,k=1}^{b_1}h_{jk}^{-1}<\alpha ,
\rho _{j}^{(1)}>\rho _{k}^{(1)},\ \forall\alpha\in\Omega ^1(M;\frak t^N).\tag4.1.21$$ From $\epsilon _{a_j}^{A_0}(A^g)=\epsilon
_{a_j}^{A_0}(A)+m_j$ and $\Pi ^{Harm^1(M)}(g^{\ast}\vartheta )=2\pi\sqrt{-1}\sum _{j=1}^{b_1}m_j\rho _j^{(1)}$ one gets $\omega
_a^{A_0}(A^g)=\omega _a^{A_0}(A)g$. According to the transition functions $\varphi _{aa^{\prime}}^{A_0}\colon U_a^{A_0}\cap
U_{a^{\prime}}^{A_0}\rightarrow\Cal G_{\ast}^{(M)}$,

$$\varphi _{aa^{\prime}}^{A_0}([A])=\hat\kappa _{(M)} ([\sum
_{j=1}^{b_1} g_{a_{j}a_{j}^{\prime}}^{\Bbb T^N}(e^{\int _M (A-A_0)\wedge\rho _j^{(n-1)}})\rho _j^{(1)}],0)\tag4.1.22$$ one concludes that the
bundle is trivializable if $H^1(M;\Bbb Z)=0$. Since the transition functions (4.1.22) are locally constant in the field space, $\Cal A^{(P)}$ is
a flat principal bundle over $\Cal M_{\ast}^{(P)}$.
\par
In the next step of the proof we want to discuss the dependence on the background connection $A_0$. However, let $A_0^{\prime}$ denote another
background connection which generates the open cover $\Cal U^{A_0^{\prime}}=\{U_a^{A_0^{\prime}}\}$ of the gauge orbit space. By passing to the
common refinement (if necessary) $\epsilon _a^{A_0^{\prime}}$ is related to $\epsilon _a^{A_0}$ by $\epsilon _a^{A_0^{\prime}}(A)=\epsilon
_a^{A_0}(A)+\hat h_{a}^{A_0,A_0^{\prime}}(A)$, where

$$\multline\hat h_{a_j}^{A_0,A_0^{\prime}}(A)=\frac{1}{2\pi\sqrt{-1}}\int
_{M}(A_0-A_0^{\prime})\wedge\rho _j^{(n-1)} +s_{a_j}( exp{\int _{M}(A-A_0)\wedge \rho _j^{(n-1)}})\\ -s_{a_j}(exp{(\int _{M}(A-A_0)\wedge \rho
_j^{(n-1)})}\cdot \exp{(\int _{M}(A-A_0^{\prime})\wedge\rho _k^{(n-1)})})\endmultline\tag4.1.23$$ is a locally constant gauge invariant function
on $\pi _{\Cal A^{(P)}}^{-1}(U_a^{A_0}\cap U_a^{A_0^{\prime}})$. Since this function $\hat h_{a_j}^{A_0,A_0^{\prime}}(A)\in\Bbb Z^N$, there
exists a map $h_a^{A_0,A_0^{\prime}}\colon U_a^{A_0}\cap U_a^{A_0^{\prime}} \rightarrow\Cal G_{\ast}^{(M)}$ which is given by

$$h_a^{A_0,A_0^{\prime}}([A])=\hat\kappa ([\sum_{j=1}^{b_1} \hat
h_{a_j}^{A_0,A_0^{\prime}}(A)\rho_j^{(1)}],G_0d_1^{\ast}(A_0-A_0^{\prime}) -G_0d_1^{\ast}(A_0-A_0^{\prime})(x_0))\tag4.1.24$$ resulting in
$\sigma _a^{A_0^{\prime}} ([A])=\sigma _a^{A_0} ([A])+(h_a([A])^{A_0,A_0^{\prime}})^{\ast}\vartheta$. This finally proves that any different
choice of the background connection gives rise to an equivalent bundle atlas of $\Cal A^{(P)}(\Cal M_{\ast}^{(P)},\pi _{\Cal A^{(P)}},\Cal
G_{\ast}^{(M)})$. This concludes the proof of theorem 4.3. \qed
\enddemo

\subheading{Remark} That the bundle of connections is non-trivial in general can be seen alternatively as follows: Since $\Cal A^{(P)}$ is
contractible, the exact homotopy sequence

$$\ldots\rightarrow\pi _k (\Cal A^{(P)} )\rightarrow\pi _k (\Cal
M_{\ast}^{(P)} )\rightarrow\pi _{k-1}(\Cal G_{\ast}^{(M)} )\rightarrow\pi _{k-1}(\Cal A^{(P)} )\rightarrow\ldots \tag4.1.25$$ implies $\pi _k
(\Cal M_{\ast}^{(P)})\cong\pi _{k-1}(\Cal G_{\ast}^{(M)} )$. If $\Cal A^{(P)} \rightarrow \Cal M_{\ast}^{(P)}$ was trivializable, then $\Cal
A^{P)}\cong\Cal M_{\ast}^{(P)}\times\Cal G_{\ast}^{(P)}$ would result in $\pi _{k-1}(\Cal G_{\ast}^{(M)})\times\pi _k (\Cal G_{\ast}^{(M)})=0$.
However, if any of the homotopy groups of $\Cal G_{\ast}^{(M)}$ does not vanish, the premise is wrong and the bundle cannot be trivializable. In
our case we have proved in proposition 4.2 that the gauge group is not connected.\par

The second important result which we are going to present is that the gauge orbit space $\Cal M_{\ast}^{(P)}$ itself admits the structure of a
bundle over a finite dimensional manifold:

\proclaim{Theorem 4.4} For each arbitrary but fixed connection $A_0\in\Cal A^{(P)}$, the manifold $\Cal M_{\ast}^{(P)}$ admits the structure of
a trivializable vector bundle over $\Bbb T^{b_1N}$ with projection $\pi _{\Cal M_{\ast}^{(P)}}^{A_0}$ and typical fiber $\Cal
N^{(M)}:=imd_2^{\ast}\otimes\frak t^N$.
\endproclaim\demo{Proof}

A bundle atlas is provided by the following local diffeomorphisms
$$\split\chi _a^{A_0} &\colon V_a\times \Cal N^{(M)}\rightarrow\Cal
M_{\ast}^{(P)}\\ \chi _{a}^{A_0}(\vec z_1,\ldots ,\vec z_{b_1},\tau )& =[A_0+2\pi\sqrt{-1}\sum _{j=1}^{b_1} s_{a_j}(\vec z_j)\rho _j^{(1)}+\tau
]\\(\chi _a^{A_0})^{-1}([A]) &=\left(\pi _{\Cal M_{\ast}^{(P)}}^{A_0}([A]), d_2^{\ast}G_2(F_A-F_{A_0})\right)\endsplit\tag4.1.26$$ On each fiber
$(\pi _{\Cal M_{\ast}^{(P)}}^{A_0})^{-1}(\vec{z}_1,\ldots ,\vec{z}_{b_1})$, there is a unique structure of a real vector space induced by the
bundle chart $\chi _a^{A_0}$, giving rise to the vector bundle structure on $\Cal M_{\ast}^{(P)}$. Here $A_0$ represents a choice of origin in
the fibers.\qed\enddemo

This concludes the analysis of the geometrical structure of the configuration space. As a consequence, the topology of the gauge orbit space is
characterized as follows:

\proclaim{Corollary 4.5}
$$\split & H^k(\Cal M_{\ast}^{(P)},\Bbb Z)=H^k(\Bbb T^{b_1N},\Bbb
Z)=\Bbb Z^{\binom {b_1N}k}\\ & \pi _1(\Cal M_{\ast}^{(P)})=\pi _0(\Cal G_{\ast}^{(M)})=\Bbb Z^{b_1N}\\ & \pi_k(\Cal M_{\ast}^{(P)})=\pi
_{k-1}(\Cal G_{\ast}^{(M)})=0\quad k\geqq 2\endsplit\tag4.1.27$$\qed\endproclaim

How does the choice of the background gauge field affect the vector bundle structure of $\Cal M_{\ast}^{(P)}$?

\proclaim{Proposition 4.6} Let $A_0$ and $A_0^{\prime}$ be two arbitrary but fixed connections. Then the fiber bundles $\Cal M_{\ast}^{(P)}@>\pi
_{\Cal M_{\ast}^{(P)}}^{A_0} >>\Bbb T^{b_1N}$ and $\Cal M_{\ast}^{(P)}@>\pi _{\Cal M_{\ast}^{(P)}}^{A_0^{\prime}}
>>\Bbb T^{b_1N}$ are isomorphic with respect to their vector bundle structures.\endproclaim

\demo{Proof} The invertible map $\Upsilon ^{A_0,A_0^{\prime}}\colon \Cal M_{\ast}^{(P)}\rightarrow\Cal M_{\ast}^{(P)}$, given by $\Upsilon
^{A_0,A_0^{\prime}}([A]):=[A+A_0^{\prime}-A_0]$ makes the following diagram of vector bundles commutative:

$$\CD \Cal M_{\ast}^{(P)} @>\Upsilon ^{A_0,A_0^{\prime}}>> \Cal M_{\ast}^{(P)} \\
@V\pi _{\Cal M_{\ast}^{(P)}}^{A_0}VV @VV \pi _{\Cal M_{\ast}^{(P)}}^{A_0^{\prime}}V \\ \Bbb T^{b_1N} @= \Bbb T^{b_1N}.\endCD\tag4.1.28$$
\qed\enddemo
\bigskip\bigskip
{\bf 4.2. The partition function, and the VEV of gauge invariant observables}\bigskip

In this section we are going to apply the results of the previous sections to calculate the partition function and the VEV of gauge invariant
functions on a closed manifold $M$. According to the defining relations in (2.2) and (2.3), this requires first of all the choice of an
appropriate gauge fixing function $S_{gf}$, which renders the gauge group volume (2.1) finite: Let $\theta =(\theta ^{\alpha})_{\alpha
=1}^{N}\in\Omega ^1(\Cal G_{\ast}^{(M)},\frak G_{\ast}^{(M)})$ denote the Maurer Cartan form on $\Cal G_{\ast}^{(M)}$. Using the Maurer Cartan
form the canonical metric (3.7) on the Lie algebra $\frak G_{\ast}^{(M)}$ can be extended to the whole gauge group. This finally generates a
left-invariant volume form $vol_{\Cal G_{\ast}^{(M)}}:=\det{(\bar\theta\theta )^{\frac{1}{2}}}\Cal Dg$ on $\Cal G_{\ast}^{(M)}$. Let us now
define a candidate for $S_{gf}$ by

$$e^{-S_{gf}(g)}=\frac{e^{-S_{gf}^{\prime}(g)}}{\int _{\Cal G_{\ast}^{(M)}}
\ vol_{\Cal G_{\ast}^{(M)}}\ e^{-S_{gf}^{\prime}}},\tag4.2.1$$ with the auxiliary gauge fixing function

$$S_{gf}^{\prime}(g)= \frac{1}{2}\Vert d^{\ast}g^{\ast}\vartheta\Vert _{\lambda}^2+
\frac{1}{2}\Vert \Pi ^{Harm^1(M)}(g^{\ast}\vartheta )\Vert _{\lambda}^2.\tag4.2.2$$ In order to prove that $S_{gf}$ indeed gives a reasonable
regularization of the gauge group we firstly recall the definition of the Riemann Theta function: Let $\Lambda$ be any symmetric complex
$r\times r$ dimensional square matrix whose imaginary part is positive definite, $u\in\Bbb C^r$ and $\alpha ,\beta\in\Bbb Z^r$ then the
$r$-dimensional Theta function is defined by

$$\Theta _{r}(u|\Lambda )=\sum\limits _{n\in\Bbb Z^r}\exp{\lbrace \pi\sqrt{-1}n
^{\dag}\cdot\Lambda\cdot n+2\pi\sqrt{-1}n^{\dag}\cdot u\rbrace },\tag4.2.3$$ where the superscript $\dag$ denotes the transpose.

\proclaim{Lemma 4.7} For the auxiliary gauge fixing function $S_{gf}^{\prime}$ in (4.2.2), the regularized volume of the gauge group yields
$$\int _{\Cal G_{\ast}^{(M)}}\ vol_{\Cal G_{\ast}^{(M)}}
e^{-S_{gf}^{\prime}}=(\det{\Delta _0|_{imd_1^{\ast}}})^{-N}\cdot\Theta _{b_1N} (0|2\pi\sqrt{-1}\ \Lambda ),\tag4.2.4$$ where $\Lambda
=\lambda\otimes h$ is the tensor product of the matrix $(\lambda _{\alpha ,\beta} )_{\alpha ,\beta =1}^N$ of coupling constants and the metric
on the harmonic 1-forms $(h _{jk} )_{j,k=1}^{b_1}$.
\endproclaim
\demo{Proof} The integral is calculated by using the isomorphism $\hat{\kappa}_{(M)}$ in (4.1.6). With respect to the fixed basis of harmonic
1-forms (4.1.16), any cohomology class $[\nu ]\in H_{\Bbb Z}^1(M;\Bbb R^N)$ gives rise to the unique harmonic representative
$\Pi^{Harm^1(M)}(\nu )=\sum _{j=1}^{b_1}m_{j}\rho _j^{(1)}$, where $m_{j}\in\Bbb Z^N$. Hence any $g\in\Cal G_{\ast}^{(M)}$ is uniquely
characterized by a pair $(\xi ,m)\in \frak G_{\ast}^{(M)}\times \Bbb Z^{b_1N}$. As a consequence the integration over $\Cal G_{\ast}^{(M)}$
means integration over $\xi$ and summation over the integers $m_{j}^{\alpha}$, where $j=1,\ldots ,b_1$ and $\alpha =1,\ldots ,N$. It is easily
shown that the integral over $\frak G_{\ast}^{(M)}$ yields the determinant of the Laplacian whereas the sum over the harmonic forms with integer
periods gives the Riemann Theta function. \qed\enddemo

At this point we would like to notice that generally all determinants of elliptic differential operators appearing in this paper are understood
in terms of the zeta function regularization [15]: For any non-negative self-adjoint elliptic operator $\Cal B$ its regularized determinant is
defined by

$$\det\Cal B =\exp{\left( -\frac{d}{ds}|_{s=0}\zeta (s|\Cal B)\right)},\tag4.2.5$$
where $\zeta (s|\Cal B)$ is the zeta-function of the operator $\Cal B$, given by

$$\zeta (s|\Cal B)=\sum _{\nu _j\neq 0}\nu _j ^{-s}=\frac{1}{\Gamma (s)}\int _0^{\infty}t^{s-1}
Tr(e^{-t\Cal B}-\Pi ^{\Cal B})dt,\tag4.2.6$$ where $\nu _j$ are the non-vanishing eigenvalues of $\Cal B$ and $\Pi ^{\Cal B}$ is the orthogonal
projector onto the kernel of $\Cal B$. Here the $\zeta$-function is analytic at the origin and possesses a meromorphic extension over $\Bbb C$.
\par
What is the geometrical meaning of the regularizing gauge fixing function $S_{gf}$? Formally $\varpi =vol_{\Cal G_{\ast}^{(M)}}\ e^{-S_{gf}}$
can be regarded as a differential form of top degree on the gauge group with integral $\int _{\Cal G_{\ast}^{(M)}}\varpi =1$. Hence $\varpi$
gives rise to a class in the cohomology with fast decrease of the gauge group. Since any other differential form $\varpi ^{\prime}$ of top
degree which integrates to one belongs to the same cohomology class than $\varpi$, any two different gauge fixing functions are ambiguous up to
an exact differential form on $\Cal G_{\ast}^{(M)}$.\par

Let us now introduce a specific partition of unity $\{p_{a}\}$ for $\Cal M_{\ast}^{(P)}$: We begin with a partition of unity
$\{\hat{p}_{a_{j\alpha}}|a_{j\alpha}\in\Bbb Z_2,\}$ on the $j\alpha$-th 1-torus $\Bbb T^1$ subordinate to the open cover $\{V_{a_{j\alpha}}\}$.
Let $q_{j\alpha}\colon\Bbb T^{b_1N}@>>>\Bbb T^1$, $q_{j\alpha}(z_{11},\ldots ,z_{j\alpha},\ldots ,z_{b_1N})=z_{j\alpha}$ be the projection onto
the $j\alpha$-th factor.  Then $\hat{p}_{a}:=\prod _{\alpha =1}^N\prod _{j=1}^{b_1}q_{j\alpha}^{\ast}\hat{p}_{a_{j\alpha}}$ induces a partition
of unity of $\Bbb T^{b_1N}$ subordinate to $\{V_{a}\}$. Finally $p_a:=\pi _{\Cal M_{\ast}^{(P)}}^{\ast}\hat p_a$ is the sought-after partition
of unity subordinate to the open cover $\Cal U^{A_0}$ of the gauge orbit space $\Cal M_{\ast}^{(P)}$. \par

Now we are prepared to display the modified global functional integral in the field space $\Cal A^{(P)}$ according to the defining relations in
(2.2) and (2.3):

\proclaim{Proposition 4.8} Let us choose the gauge fixing function $S_{gf}$ (4.2.1). The partition function for the abelian gauge theory on a
closed manifold $M$ with the classical action (3.6) is given by

$$Z_{A_0}^{(P)}=
\int _{\Cal A^{(P)}} vol_{\Cal A^{(P)}}\ \Cal F(A)\cdot e^{-\frac{1}{2}(\|F_{A}\|_{\lambda}^2+\|d^{\ast}(A-A_0)\|_{\lambda}^2)},\tag4.2.7$$
where

$$\Cal F(A)= (\det{\Delta _0|_{imd_0^{\ast}}})^{N}\cdot
\Theta _{b_1N}(0|2\pi\sqrt{-1}\Lambda )^{-1}\sum _{a\in\Bbb Z_2^{b_1N}} (\pi _{\Cal A^{(P)}}^{\ast}p_a) e^{-2\pi ^2\sum\limits _{\alpha ,\beta
=1}^{N}\sum \limits _{j,k=1}^{b_1} \lambda _{\alpha\beta}h_{jk}\epsilon _{a_{j\alpha}}^{A_0}(A)\epsilon _{a_{k\beta}}^{A_0}(A)},\tag4.2.8$$ with
the multi-index $a=(a_{11},\ldots ,a_{b_1N})\in\Bbb Z_2^{b_1N}$.
\endproclaim

\demo{Proof} The formula for the partition function can be verified directly by using the bundle coordinates (4.1.20) and the gauge fixing
function (4.2.2). \qed\enddemo

Unlike the conventional Faddeev-Popov result for the partition function there has appeared an additional contribution in $\Cal F(A)$ caused by
the non-triviality of the bundle of connections and the non-compactness of the gauge group. Since $\Cal F(A)$ is non-vanishing and positive, the
interpretation of the integrand as a probability density remains valid. In the case of simply connected manifolds $M$, where the Gribov problem
is absent, eq. (4.2.7) does reproduce exactly the conventional Faddeev-Popov formula for the abelian gauge theory in the Lorentz gauge. In fact,
$\Cal F(A)$ reduces to the field independent multiplicative constant $(\det{\Delta _0|_{imd_1^{\ast}}})^{N}$ which according to lemma 4.7 is
nothing but the inverse of the regularized volume of the subgroup of infinitesimal gauge transformations. \par

What is the effect of $\Cal F(A)$ in the topologically non-trivial case where it modifies the gauge fixed action? Since the bundle of
connections is flat, the functional integral (4.2.7) can be decomposed on each $\pi _{\Cal A^{(P)}}^{-1}(U_a)\subset\Cal A^{(P)}$ into a
disjoint union of open sets, each of them diffeomorphic to the product of the gauge fixing submanifold $\{A|\omega _a^{A_0}(A)=1\}$ and a
sufficiently small open set of those gauge transformations which are connected to the identity. On each of these slices the functionals
$\epsilon _{a_{j\alpha}}^{A_0}$ are constant giving rise to a regularization of gauge transformations not connected to the identity. Moreover,
the partition of unity $p_a$ is constant on each $\pi _{\Cal A^{(P)}}^{-1}(U_a)$. We will see in the sequel that the VEV\rq s of gauge invariant
observables are not affected by the explicit form of the gauge fixing function $S_{gf}$. \par

In the next step we aim to find an explicit expression for the partition function in (4.2.7). The strategy is to split the gauge fields
$A\in\Cal A^{(P)}$ into components according to the bundle structures which were described by the theorems 4.2 and 4.3 and then to split the
integration over $\Cal A^{(P)}$ into an integration over the base manifold $\Bbb T^{b_1N}$ and an integration over the fiber $\Cal
G_{\ast}^{(M)}\times\Cal N^{(M)}$. \par

We shall begin with the decomposition of the volume form $vol_{\Cal A^{(P)}}$. Let us define the local diffeomorphisms $\psi _a^{A_0}=\varphi
_a^{A_0}\circ (\chi _a^{A_0}\times\Bbb I)\colon V_a\times \Cal N^{(M)}\times\Cal G_{\ast}^{(M)}\rightarrow (\pi _{\Cal
M_{\ast}^{(P)}}\circ\pi_{\Cal A^{(P)}})^{-1}(V_a)$. Then the differential of $\psi _a^{A_0}$ can be easily calculated to yield

$$T_{(\vec{z}_1,\ldots ,\vec{z}_{b_1},\tau ,g)}\psi _a^{A_0}(\vec{w}_1,\ldots ,
\vec{w}_{b_1},u ,\Cal Y)=\sum _{\alpha =1}^N\sum _{j=1}^{b_1}\vartheta _{z_{j\alpha}}^{\Bbb T^1}(w_{j\alpha})\rho _j^{(1)}+u+d\theta _g(\Cal
Y),\tag4.2.9$$ where $\vartheta ^{\Bbb T^1}$ is the Maurer Cartan form on $\Bbb T^1$, $\vec{z}_j=(z_{j1},\ldots z_{jN})\in\Bbb T^N$ for
$j=1,\ldots b_1$ and $\vec{w}_j=(w_{j1},\ldots ,w_{jN})\in T_{\vec{z}_j}\Bbb T^{N}$, $u\in T_{\tau}\Cal N^{(M)}$ and $\Cal Y\in T_g\Cal
G_{\ast}^{(M)}$. The metric (3.7) can be recasted into

$$\split &((\psi _a^{A_0})^{\ast}<,>_{\lambda})_{(\vec{z}_1,\ldots ,\vec{z}_{b_1},\tau ,g)}
\left( (\vec{w}_1^{1},\ldots , \vec{w}_{b_1}^{1},u ^{1},\Cal Y^{1}),(\vec{w}_1^{2},\ldots , \vec{w}_{b_1}^{2},u ^{2},\Cal Y^{2})\right)=\\ &
=\sum \Sb \alpha =1\\ \beta =1\endSb ^N\sum \Sb j=1\\ k=1\endSb ^{b_1}\lambda _{\alpha\beta}\overline{\vartheta _{z_{j\alpha}}^{\Bbb
T^1}(w_{j\alpha}^{1})}\vartheta _{z_{k\beta}}^{\Bbb T^1}(w_{k\beta}^{2})h_{jk}+<u^{1},u^{2}>_{\lambda}+<d\theta _g (\Cal Y^{1}),d\theta _g(\Cal
Y^{2})>_{\lambda}.\endsplit\tag4.2.10$$ Formally (4.2.10) can be equivalently rewritten into the following matrix form

$$((\psi _a^{A_0})^{\ast}<,>_{\lambda})_{(\vec{z}_1,\ldots ,\vec{z}_{b_1},\tau ,g)}
=\pmatrix h_{jk}\lambda _{\alpha\beta} \overline{\vartheta
_{z_{j\alpha}}^{\Bbb T^1}}\vartheta _{z_{k\beta}}^{\Bbb T^1} & 0 & 0\\ 0 & \lambda _{\alpha\beta} & 0\\
0 & 0 & \lambda _{\alpha\beta}\bar\theta _g^{\alpha}\Delta _0\theta _g^{\beta}\endpmatrix\tag4.2.11$$ with $\bar\vartheta ^{\Bbb T^1}$ and
$\bar\theta$ denoting the complex conjugates of the Maurer Cartan forms on $\Bbb T^1$ and $\Cal G_{\ast}^{(M)}$ respectively. Each component in
(4.2.11) displays the induced metric on the corresponding space. In terms of the local trivialization the volume form becomes

$$(\psi _a^{A_0})^{\ast}vol_{\Cal A^{(P)}}= (\det h)^{N/2}\det{(\Delta _0|_{imd_1^{\ast}})}^{N/2}\ vol_{\Bbb
T^{b_1N}}\vert _{V_a}\wedge vol_{\Cal N^{(M)}}\wedge vol_{\Cal G_{\ast}^{(M)}},\tag4.2.12$$ where $vol_{\Bbb
T^{b_1N}}=(\sqrt{-1})^{-b_1N}q_{11}^{\ast}\vartheta ^{\Bbb T^1}\wedge\ldots \wedge q_{b_1N}^{\ast}\vartheta ^{\Bbb T^1}$ is the induced volume
form on $\Bbb T^{b_1N}$, which in (4.2.12) is restricted to the patch $V_a$. The volume form $vol_{\Cal N^{(M)}}$ is induced by the flat metric
on $\Cal N^{(M)}$ and can be formally written as $vol_{\Cal N^{(M)}}=\Cal D\tau$.

\proclaim{Lemma 4.9} The background connection $A_0\in\Cal A^{(P)}$ can be chosen to satisfy the classical equation of motion,
$d_2^{\ast}F_{A_0}=0$.\endproclaim \demo{Proof} Given any background gauge field $A_0^{\prime}$ the modified background connection
$A_0=A_0^{\prime}-G_1d_2^{\ast}F_{A_0^{\prime}}$ satisfies the requested equation. \qed\enddemo

\proclaim{Proposition 4.10} There exists a globally defined density $\hat\Xi ^{(P)}$ on the direct product $\Bbb T^{b_1N}\times \Cal
N^{(M)}\times \Cal G_{\ast}^{(M)}$ so that $(\psi _a^{A_0})^{\ast}\Xi _a^{(P)}=i_{V_a}^{\ast}\hat\Xi ^{(P)}$, where $i_{V_a}\colon
V_a\hookrightarrow\Bbb T^{b_1N}$ is the restriction to $V_a$.
\endproclaim

\demo{Proof} Using (2.2) and (4.2.12) one verifies by a direct calculation that

$$\hat\Xi ^{(P)} =(\det h)^{N/2}(\det{\Delta _0|_{imd_1^{\ast}}})^{N/2}
\ vol_{\Bbb T^{b_1N}}\wedge vol_{\Cal N^{(M)}}\wedge vol_{\Cal G_{\ast}^{(M)}} e^{-\frac{1}{2}(\Vert F_{A_0}\Vert _{\lambda}^2+<\tau ,\Delta
_1|_{\Cal N^{(M)}}\tau >_{\lambda})-S_{gf}}.\tag4.2.13$$ gives the global density with the required property. \qed\enddemo

Geometrically the existence of the global density $\hat\Xi ^{(P)}$ is related to the fact that the bundle of connections is flat in the abelian
gauge theory. Since the bundle of connections is not flat in the non-abelian Yang-Mills theory [3,4], the corresponding local densities cannot
be extended to a global form without use of a partition of unity.
\par
Let $f$ be a gauge invariant observable, i.e. a (real-valued) function on $\Cal A^{(P)}$, satisfying $f(A^g)=f(A)$. In the following we will
denote the induced function on $\Cal M_{\ast}^{(P)}$ with the same symbol. Then $\hat{f}:=(\chi _a^{A_0})^{\ast}f$ is a globally defined
function on $\Bbb T^{b_1N}\times \Cal N^{(M)}$.\par

Let $e_{(m_{11},\ldots ,m_{b_1N})}(z_{11},\ldots ,z_{b_1N}):=z_{11}^{m_{11}}\cdots z_{b_1N}^{m_{b_1N}}$ be an orthonormal basis of $L^2(\Bbb
T^{b_1N};\Bbb C)$ with respect to the inner product $\ll f_1,f_2 \gg:=\frac{1}{(2\pi )^{b_1N}}\int _{\Bbb T^{b_1N}}vol_{\Bbb
T^{b_1N}}\bar{f}_1f_2$, where $z_{j\alpha}\in\Bbb T^1$ and $m_{j\alpha}\in\Bbb Z$. Then $\hat{f}(.,A_0+\tau )$ can be rewritten in terms of a
Fourier series expansion on $\Bbb T^{b_1N}$ as

$$\hat{f}(.,A_0+\tau )=\sum _{m_{11}\in\Bbb Z}\cdots\sum _{m_{b_1N}\in\Bbb Z}
\hat{f}_{(m_{11},\ldots ,m_{b_1N})}(A_0+\tau )\ e_{(m_{11},\ldots ,m_{b_1N})},\tag4.2.14$$ with Fourier coefficients

$$\multline\hat{f}_{(m_{11},\ldots ,m_{b_1N})}(A_0+\tau ):=\ll e_{(m_{11},\ldots ,m_{b_1N})},\hat{f}(.,A_0+\tau ))\gg=\\
=\int\limits _{0}^{1}\cdots\int\limits _{0}^{1}dt_{11}\ldots dt_{b_1N}\hat{f}(e^{2\pi\sqrt{-1}t_{11}},\ldots ,e^{2\pi\sqrt{-1}t_{b_1N}},A_0+\tau
)e^{-2\pi\sqrt{-1}\sum\limits _{\alpha =1}^N\sum\limits _{j=1}^{b_1}m_{j\alpha}t_{j\alpha}}.\endmultline\tag4.2.15$$ Using (4.2.13) and (4.2.14)
we get

$$\split I^{(P)}(f)= &\sum\limits _{a\in\Bbb Z_2^{b_1N}}\ \int\limits _{V_a\times \Cal N^{(M)}\times\Cal G_{\ast}^{(M)}}
\ (pr_{\Bbb T^{b_1N}}^{\ast}\hat p_a)\cdot i_{V_a}^{\ast}\hat\Xi ^{(P)} \cdot pr_{\Cal G_{\ast}^{(M)}}^{\ast}(\chi _a^{A_0})^{\ast}f =\\ =
&(2\pi )^{b_1N}(\det h)^{N/2} (\det{\Delta _0|_{imd_1^{\ast}}})^{N/2}e^{-\frac{1}{2}||F_{A_0}||_{\lambda}^2}\int\limits _{\Cal N^{(M)}}\Cal
D\tau\ \hat{f}_{(0,\ldots ,0)}(A_0+\tau )\ e^{-\frac{1}{2}<\tau ,\Delta _1|_{\Cal N^{(M)}}\tau
>_{\lambda}} ,\endsplit\tag4.2.16$$ where $pr_{\Bbb T^{b_1N}}$ and $pr_{\Cal G_{\ast}^{(M)}}$ are the projections onto the first and third factor
in $\Bbb T^{b_1N}\times \Cal N^{(M)}\times\Cal G_{\ast}^{(M)}$, respectively. Because of the gauge invariance, the integral $I^{(P)}(f)$ is
independent of the explicit form of the partition of unity. Using that $\det{(\Delta _{1}|_{imd_{0}})}=\det{(\Delta _0|_{imd_1^{\ast}})}$ one
gets from (4.2.16):

\proclaim{Proposition 4.11} For any fixed component $\Cal A^{(P)}$ the following holds:\roster\item The partition function $Z_{A_0}^{(P)}$ is
given by
$$Z_{A_0}^{(P)} =(2\pi )^{b_1N}(\det
h)^{N/2} (\det{\Delta _0|_{imd_1^{\ast}}})^{N} (\det{\Delta _1|_{Harm^1(M)^{\perp}}})^{-N/2}
e^{-\frac{1}{2}\|F_{A_0}\|_{\lambda}^2}.\tag4.2.17$$

\item The VEV of any gauge invariant function $f$ reads

$$<f>_{(P)}=(\det{\Delta _0|_{imd_1^{\ast}}})^{-N/2}\cdot
(\det{\Delta _1|_{Harm^1(M)^{\perp}}})^{N/2}\int\limits _{\Cal N^{(M)}}\ \Cal D\tau\hat{f}_{(0,\ldots ,0)}(A_0+\tau )\cdot e^{-\frac{1}{2}<\tau
,\Delta _1|_{\Cal N^{(M)}}\tau
>_{\lambda}}.\tag4.2.18$$
\endroster\qed\endproclaim

Let us take a simply connected manifold $M$ and consider the trivial $\Bbb T^N$-bundle $P\cong M\times \Bbb T^N$. Then one can choose $A_0=0$.
In that case $\Cal M_{\ast}^{(P)}\cong \Cal N^{(M)}$ implying that the inequivalent gauge fields are in one-to-one correspondence with the space
of transversal fields. For $N=1$ the partition function reduces to

$$Z_{A_0=0}^{(M\times \Bbb T^N)}=(\det{\Delta _0|_{imd_1^{\ast}}})(\det{\Delta
_1})^{-1/2},\tag4.2.19$$ which is the well-known covariant expression for the quantized Maxwell theory (see e.g. [15]). \par

Let us reflect on the difference between the modified functional integral and the Faddeev-Popov procedure once again. Usually it is called that
a convergent factor is introduced by the gauge fixing in a way that does not affect the VEV of any gauge invariant observable. However, if
Gribov ambiguities are present the conventional gauge fixed partition function in the Lorentz gauge

$$\int _{\Cal A^{(P)}}vol _{\Cal A^{(P)}}\
e^{-S_{inv}(A)-\frac{1}{2}||d^{\ast}(A-A_0)||^2},\tag4.2.20$$ would never yield a finite functional integral: In fact, rewriting (4.2.20) in
terms of the local coordinates $\{\psi _a^{A_0}\}$ and using proposition 4.2 the following divergent integral
$$\int _{\Cal G_{\ast}^M}\
vol_{\Cal G_{\ast}^M}\ e^{-\frac{1}{2}||d^{\ast}g^{\ast}\vartheta||^2}\tag4.2.21$$ appears as a factor in the total functional integral. This
infinite factor is a consequence of the fact that the conventional gauge fixing term $\frac{1}{2}||d^{\ast}(A-A_0)||^2$ does not damp the gauge
transformations which are not connected to the unity. Nevertheless this term is sufficient to regularize the subgroup of infinitesimal gauge
transformations.
\par

\subheading{The partition function on $\Cal A^{(M)}$} The principal $\Bbb T^N$-bundles $P$ are labelled by their Chern-class
$c=(c^{\alpha})_{\alpha =1}^{N}\in H^2(M;\Bbb Z^N)$. Since the cohomology of $M$ is finitely generated, $c$ takes the form

$$c^{\alpha}=\sum _{j=1}^{b_2}m_{j\alpha}c_{j}^{(2)}+
\sum _{k=1}^{r}t_{k\alpha}y_{k}^{(2)},\tag4.2.22$$ where $(c_{j}^{(2)})_{j=1}^{b_2}$ denotes a Betti basis of $H^2(M;\Bbb Z)$, $b_2=dim
H^2(M;\Bbb R)$ and $m_{j\alpha}\in\Bbb Z$ for $j=1,\ldots ,b_2$, $\alpha =1,\ldots , N$. On the other hand $TorH^2(M;\Bbb Z)$ is generated by a
basis $(y_{k}^{(2)})_{k=1}^{r}$ with torsion coefficients $l_k$, i.e. $l_ky_{k}^{(2)}=0$ and $t_{k\alpha}\in\Bbb Z_{l_k}$ for $k=1,\ldots ,r$,
$\alpha =1,\ldots ,N$. According to lemma (4.9), $F_{A_0}\in Harm_{\Bbb Z}^2(M;\Bbb R)\otimes\frak t^N$. Let $\rho _j^{(2)}\in Harm_{\Bbb
Z}^2(M;\Bbb R)$, for $j=1,\ldots ,b_2$, be a basis of harmonic two forms on $M$ with integer periods, and let $h_{jk}^{(2)}=<\rho _j^{(2)},\rho
_k^{(2)}>$ denote the induced metric on $Harm^2(M;\Bbb R)$. Then the field strength can be rewritten into

$$F_{A_0}^{\alpha}=2\pi\sqrt{-1}\sum _{k=1}^{b_2}m_{k\alpha}\rho
_k^{(2)},\quad m_{k\alpha}=\sum _{j=1}^{b_2}(h_{jk}^{(2)})^{-1}<F_{A_0}^{\alpha},\rho _{j}^{(2)}> \in\Bbb Z.\tag4.2.23$$ Let us define the
tensor product $\Lambda ^{(2)}:=\lambda\otimes h^{(2)}$, then the sum over the equivalence classes of principal bundles $(P)$ in (2.6) can be
split into a sum over the free part and the torsion part of $H^2(M;\Bbb Z^N)$. Hence we obtain from (2.6):

\proclaim{Proposition 4.12} The partition function $Z$ for the abelian gauge theory on the total configuration space $\Cal A^{(M)}$ is given by

$$\split Z = &(2\pi )^{b_1N}(\det
h)^{N/2} (\det{\Delta _0|_{imd_1^{\ast}}})^{N} (\det{\Delta _1|_{Harm^1(M)^{\perp}}})^{-N/2}\\ &\times\Theta _{Nb_2}(0|2\pi\sqrt{-1}\Lambda
^{(2)}) ord(TorH^2(M;\Bbb Z^N)),\endsplit\tag4.2.24$$ where $ord(TorH^2(M;\Bbb Z^N))$ is the order of the finite torsion subgroup of $H^2(M;\Bbb
Z^N)$.
\par\qed\endproclaim

The partition function $Z$ does not depend on the choice of the basis $\{\rho _j^{(1)}\}$ and $\{\rho _j^{(2)}\}$ because any other basis of
$Harm_{\Bbb Z}^{k}(M;\Bbb R)$ ($k=1,2$) is connected by a unimodular transformation, under which both $\det h$ and the $\Theta$-function remain
invariant.
\bigskip
\subheading{The correlation functions for the field strength} As an example we shall apply our results to the determination of the VEV of a
polynomial in the field strength $F_A$. This VEV is understood in the following distributional sense

$$\Cal W_q^{(P)}(\eta _1\ldots ,\eta _q):=<\prod _{j=1}^q<F_A,\eta
_j>>_{P},\tag4.2.25$$ for $\eta _1,\ldots ,\eta _q\in \Omega ^2(M;\frak t^N)$. To calculate (4.2.25) let us consider the gauge invariant
observable $f(A,\hat J)=e^{<F_A,\hat{J}>}$ with source $\hat{J}\in\Omega ^2(M;\frak t^N)$. Its VEV follows from (4.2.18)

$$\split <e^{<F_A,\hat{J}>}>_{P}= &(2\pi )^{b_1N}(\det h)^{N/2}
(\det{\Delta _0|_{imd_1^{\ast}}})^{N}(\det{\Delta
_1|_{Harm^{1}(M)^{\perp} }})^{-N/2}\\
&\times e^{-\frac{(2\pi )^2}{2}\sum\limits _{\alpha ,\beta =1}^N\sum\limits _{j,k =1}^{b_2}\lambda
_{\alpha\beta}h_{jk}^{(2)}m_{j\alpha}m_{k\beta}+2\pi\sum\limits _{\alpha =1}^N\sum\limits _{j=1}^{b_2}
m_{j\alpha}<\sqrt{-1}\rho _{j}^{(2)},\hat{J^{\alpha}}>}\\
&\times e^{\frac{1}{2}\sum\limits _{\alpha ,\beta =1}^N (\lambda ^{-1})_{\alpha\beta}<\hat{J}^{\alpha},(\Pi
^{Harm^2(M)^{\perp}}-d_{3}^{\ast}G_3d_2)\hat{J}^{\beta}>}.\endsplit\tag4.2.26$$ Finally $\Cal W_q$ can be obtained from (4.2.26) by
differentiation, namely

$$\Cal W_q^{(P)}(\eta _1\ldots ,\eta _q) =\frac{\partial ^q}{\partial t_1\cdots\partial t_q}
\vert _{t_1=\ldots =t_q=0} \ <e^{<F_A,\sum _{i=1}^qt_i\eta _i>}>_{(P)}.\tag4.2.27$$ Let us now introduce the following abbreviations
$$\align \mu_{ij} &:=<\eta _i,(\Pi
^{Harm^2(M)^{\perp}}-d_{3}^{\ast}G_3d_2)\eta _j>_{\lambda ^{-1}}\\
\nu _i &:= <F_{A_0},\eta _i>,\tag4.2.28\endalign$$ and let $\varsigma$ denote a permutation of the indices $\{1,\ldots ,q\}$ then a lengthy
calculation gives the following:

\proclaim{Proposition 4.13} For any fixed component $\Cal A^{(P)}$, the VEVs for the field strength of degree $q$ are given by
\par
1) $q=2k$
$$\split \Cal W_{2k}^{(P)} (\eta _1\ldots ,\eta _{2k}) &=\frac{1}{(2k)!}
\sum _{\varsigma} \nu _{\varsigma (1)}\cdots\nu _{\varsigma (2k)}\\
& +\sum _{l=1}^{k-1}\frac{1}{2^ll!(2(k-l))!} \sum _{\varsigma} \mu _{\varsigma (1)\varsigma (2)}\cdots\mu _{\varsigma (2l-1)\varsigma
(2l)}\nu _{\varsigma (2l+1)}\cdots\nu _{\varsigma (2k)}\\
&+\frac{1}{2^kk!}\sum _{\varsigma}\mu _{\varsigma (1)\varsigma (2)}\cdots\mu _{\varsigma (2k-1)\varsigma (2k)}
\endsplit\tag4.2.29$$

2) For $q=2k+1$
$$\split \Cal W_{2k+1}^{(P)} (\eta _1\ldots &,\eta _{2k+1})=\frac{1}{(2k+1)!}
\sum _{\varsigma}\nu _{\varsigma (1)}\cdots\nu _{\varsigma (2k+1)}\\
 &+\sum _{l=1}^{k}\frac{1}{2^ll!(2(k-l)+1)!}\sum _{\varsigma} \mu
_{\varsigma (1)\varsigma (2)}\cdots\mu _{\varsigma (2l-1)\varsigma (2l)}\nu _{\varsigma (2l+1)}\cdots\nu _{\varsigma
(2k+1)}\endsplit\tag4.2.30$$ \qed
\endproclaim
By construction (4.2.29) and (4.2.30) are independent of the actual choice for the gauge fixing function.
\bigskip
\bigskip
{\bf 4.3. The Green\rq s functions for the gauge fields}
\bigskip
Let $J\in \Omega ^1(M;\frak t^N)$ be the source for the gauge fields then we define the generating functional by

$$Z_{A_0}^{(P)}[J]=\int _{\Cal A^{(P)}}\sum _{a\in\Bbb Z_2^{b_1N}}
(\pi _{\Cal A^{(P)}}^{\ast}p_a)\cdot\Xi _a^{(P)}\cdot e^{<J,A-A_0>},\tag4.3.1$$ which is a generalization of the definition for the generating
functional around classical solutions [31]. From (4.3.1) the $q$-point Green\rq s functions $\Cal S_q$ are constructed as follows:

$$\Cal S_q^{(P)}(v_1\ldots ,v_q) :=\frac{\partial ^q}{\partial t_1\cdots\partial t_q}
\vert _{t_1=\ldots =t_q=0} \ \frac{Z_{A_0}^{(P)}[\sum _{i=1}^qt_iv_i;A_0]}{Z_{A_0}^{(P)}[0]},\tag4.3.2$$ for $v_1,\ldots ,v_q\in \Omega
^1(M;\frak t^N)$. In order to find an explicit expression for the Green\rq s functions we shall recast $Z^{(P)}[J;A_0]$ in terms of the local
trivialization $\{\psi _a^{A_0}\}$. Let $\Pi ^{imd_0}:=d_0G_0d_1^{\ast}$ be the projector onto the space of exact one-forms on $M$ then the
integration over the gauge group yields:

\proclaim{Lemma 4.13} Let us choose the auxiliary gauge fixing function $S_{gf}^{\prime}$ in (4.2.2). Then the integration over the gauge group
gives

$$\int _{\Cal G_{\ast}^{(M)}} vol_{\Cal G_{\ast}^{(M)}}
e^{-S_{gf}^{\prime}(g)+<J,g^{\ast}\vartheta >} =(\det{\Delta _0|_{imd_1^{\ast}}})^{-N}\cdot e^{1/2<\Pi ^{imd_0}(J),G_1\Pi
^{imd_0}(J)>}\cdot\Theta _{b_1N}(K(J)|2\pi\sqrt{-1}\Lambda ),\tag4.3.3$$ where $K_{j}^{\alpha}(J)=-\sqrt{-1}<J^{\alpha},\sqrt{-1}\rho _j^{(1)}
>$ with $j=1,\ldots ,b_1$ and $\alpha =1,\ldots ,N$ is regarded as $b_1N$-dimensional complex vector $K(J)$. \qed\endproclaim

Using (4.2.9) and (4.3.3) one obtains for the generating functional in (4.3.1)

$$\split Z_{A_0}^{(P)}[J]&=(\det h)^{N/2}(\det{\Delta_0|_{imd_1^{\ast}}})^{N}\cdot (\det{\Delta_1|_{Harm^1(M)^{\perp}}})
^{-N/2}\cdot e^{-\frac{1}{2}\|F_{A_0}\|_{\lambda }^2}\cdot e^{\frac{1}{2}<J,G_1J>_{\lambda ^{-1}}}\\ &\times \Theta
_{b_1N}(K(J)|2\pi\sqrt{-1}\Lambda )\cdot \Theta _{b_1N}(0|2\pi\sqrt{-1}\Lambda )^{-1}\\ &\times\int _{\Bbb T^{b_1N}} vol_{\Bbb T^{b_1N}} \sum
_{a_{11}=1}^2\ldots\sum _{a_{b_1N}=1}^2 q_{11}^{\ast}\hat p_{a_{11}}\cdots q_{b_1N}^{\ast}\hat p_{a_{b_1N}}e^{2\pi\sum _{\alpha =1}^N\sum
_{j=1}^{b_1}q_{j\alpha}^{\ast}s_{a_{j\alpha}}<J^{\alpha},\sqrt{-1}\rho _j^{(1)}>}.\endsplit\tag4.3.4$$ Defining the two field independent
factors

$$\split &\varepsilon _{j\alpha}^{(1)}:=\sum\limits _{a_{j\alpha}=1}^2\int _{V_{a_{j\alpha}}}vol_{\Bbb T^1}|_{V_{a_{j\alpha}}}
\hat p_{a_{j\alpha}}s_{a_{j\alpha}} \\
&\varepsilon _{j\alpha ,k\beta}^{(2)}:= \int _{\Bbb T^{b_1N}} vol_{\Bbb T^{b_1N}} \sum _{a_{11}=1}^2\ldots\sum _{a_{b_1N}=1}^2 q_{11}^{\ast}\hat
p_{a_{11}}\cdots q_{b_1N}^{\ast}\hat p_{a_{b_1N}}\cdot q_{j\alpha}^{\ast}s_{a_{j\alpha}}\cdot
q_{k\beta}^{\ast}s_{a_{k\beta}}\endsplit\tag4.3.5$$ one finally ends up with the following result:

\proclaim{Proposition 4.14} For any fixed component $\Cal A^{(P)}$ and chosen gauge fixing function (4.2.1), the Green\rq s functions are given
by
\par 1) One-point function:

$$\Cal S_1^{(P)}(v ) =\sum _{\alpha =1}^N\sum
_{j=1}^{b_1}\varepsilon _{j\alpha}^{(1)} <v^{\alpha},\sqrt{-1}\rho _j^{(1)}>+ \frac{d}{dt}|_{t=0}\ln{\Theta _{b_1N}(K(tv)|2\pi\sqrt{-1}\Lambda
)}.\tag4.3.6$$

2) Two-point function:

$$\split &\Cal S_2^{(P)}(v_1,v_2) = <v_1,
G_1v_2>_{\lambda ^{-1}} \\ &+\sum _{\alpha =1}^N\sum _{j=1}^{b_1}\varepsilon _{j\alpha}^{(1)}<v_{1}^{\alpha},\sqrt{-1}\rho
_j^{(1)}>(\frac{d}{dt}|_{t=0}\ln{\Theta _{b_1N}(K(tv_2
)|2\pi\sqrt{-1}\Lambda )}) \\
&+ \sum _{\alpha =1}^N\sum _{j=1}^{b_1}\varepsilon _{j\alpha}^{(1)}<v _{2}^{\alpha},\sqrt{-1}\rho _j^{(1)}>(\frac{d}{dt}|_{t=0}\ln{\Theta
_{b_1N}(K(tv_1 )|2\pi\sqrt{-1}\Lambda )})\\ &+ (2\pi )^{2-b_1N}\sum _{\alpha ,\beta =1}^N\sum _{j,k=1}^{b_1}\varepsilon _{j\alpha
,k\beta}^{(2)}<v_{1}^{\alpha},\sqrt{-1}\rho _j^{(1)}><v
_{2}^{\beta},\sqrt{-1}\rho _k^{(1)}>\\
&+\Theta _{b_1N}(0|2\pi\sqrt{-1}\Lambda )^{-1}\frac{\partial ^2}{\partial t_1\partial t_2} \vert _{t_1=t_2=0}\Theta _{b_1N}(K(\sum
_{l=1}^2t_lv_l)|2\pi\sqrt{-1}\Lambda )
\endsplit\tag4.3.7$$
\qed\endproclaim

On manifolds with vanishing first Betti number, the equations (4.3.6) and (4.3.7) yield the well-known result: The one-point function vanishes
and the two-point function reduces to the Greens operator $G_1$.\par

Before closing this section we want to display explicit results for the factors in (4.3.5). In the following local coordinate system of $\Bbb
T^1$
$$\align v_{1}\colon V_{1} \rightarrow (0,1 )\quad v_{1}^{-1}(t) &
=(\cos{2\pi t},\sin{2\pi t})\\  v_{2}\colon V_{2} \rightarrow (-\frac{1}{2} ,\frac{1}{2} )\quad v_{2}^{-1}(t) & =(\cos{2\pi t},\sin{2\pi
t}),\tag4.3.8\endalign$$ a partition of unity subordinate to $V_{j\alpha}\subset\Bbb T^1$ (see subsection 4.2) can be given by $\hat p_1(t)=\sin
^2(\pi t)$ and $\hat p_2(t)=\cos ^2(\pi t)$. Then

$$\split\varepsilon _{j\alpha}^{(1)} &=\frac{3\pi}{2}\\
\varepsilon _{j\alpha ,k\beta}^{(2)} &= \cases (2\pi )^{b_1N-1}(\frac{17\pi}{12}-\frac{1}{\pi}), &\text{for $j=k$ and $\alpha =\beta$}\\ (2\pi
)^{b_1N-2}(\frac{3\pi}{2})^2, &\text{for $j\neq k$ or $\alpha\neq\beta$ or both}.\endcases
\endsplit\tag4.3.9$$ Any other local section $s_a^{\prime}$ of (4.1.7) is connected with the section $s_a$, which was defined in (4.1.10), by
$s_a^{\prime}:=(s_{a_{11}}^{\prime},\ldots ,s_{a_{b_1N}}^{\prime})=(s_{a_{11}}+m_{a_{11}}^{11},\ldots ,s_{a_{b_1N}}+m_{a_{b_1N}}^{b_1N})$ with
$m_{a_{j\alpha}}^{j\alpha}\in\Bbb Z$ for $j=1,\ldots b_1$ and $\alpha =1,\ldots ,N$. In terms of these new sections, the factors in (4.3.5)
become
$$\split\varepsilon
_{j\alpha}^{\prime (1)} &=\pi (m_{1}^{j\alpha}+ m_{2}^{j\alpha}+\frac{3}{2}) \\
\varepsilon _{j\alpha ,k\beta}^{\prime (2)} &= \cases (2\pi )\left(\frac{17\pi}{12}-\frac{1}{\pi}+\pi
(m_{1}^{j\alpha}(m_{1}^{j\alpha}+1)+m_{2}^{j\alpha}(m_{2}^{j\alpha}+2))\right),\\
\quad\quad\quad\quad\quad\quad\text{for $j=k$ and $\alpha =\beta$}
\\ \pi
^2(m_{1}^{j\alpha}+m_{2}^{j\alpha}+\frac{3}{2})(m_1^{k\beta}+m_{2}^{k\beta}+\frac{3}{2}),
\\ \quad\quad\quad\quad\quad\quad\text{for $j\neq k$ or $\alpha\neq\beta$ or
both}.\endcases\endsplit\tag4.3.10$$ Thus it is not possible to arrange a local trivialization of $\Cal A^{(P)}$ in such a way that the
additional contributions in the Green\rq s functions would vanish. The novel feature of the modified functional integral is that the existence
of Gribov ambiguities affects the $q$-th point Green\rq s functions of the abelian gauge theory. In particular, the vacuum expectation value of
the gauge field $A$ does not vanish in general on non-simply connected manifolds.
\bigskip
\bigskip
{\bf 5. Abelian gauge theories on manifolds with boundary}\bigskip

In this chapter we want to address the construction of the modified functional integral for the abelian gauge theory with the classical action
(3.6) over a manifold without non-empty boundary. The functional integral on such manifolds requires boundary conditions to be imposed on the
fields on $\partial M$. In consequence the functional integral will become a functional of the fields on the boundary.\bigskip\bigskip {\bf 5.1.
The geometry of gauge fields}
\bigskip
Let $M$ denote a $n$-dimensional compact, connected and oriented manifold with a non-empty boundary $\partial M$, where $i_{\partial
M}\colon\partial M\rightarrow M$ is the inclusion. We choose an arbitrary but fixed principal $\Bbb T^N$-bundle $Q(\partial M,\pi _Q,\Bbb T^N)$
over $\partial M$. Let us consider the set $\frak P_Q[M;\Bbb T^N]$ of those principal $\Bbb T^N$-bundles $P(M,\pi _P,\Bbb T^N)$ over $M$ which,
if pulled back to the boundary $\partial M$, are isomorphic to $Q$ via a map $\phi$ such that the following diagram of bundles commutes:
$$\CD Q @>\phi >>\partial P:=i_{\partial M}^{\ast}P @>\hat i_{\partial M}>> P
\\@V\pi _Q VV @V\pi _{\partial P} VV @V\pi _P VV \\ \partial M @> id >>
\partial M @>i_{\partial
M}>> M.\endCD\tag5.1.1$$ Boundary conditions on the gauge fields are imposed by choosing a fixed but arbitrary connection $B\in\Cal A^{(Q)}$.
For a given $P\in\frak P_Q[M;\Bbb T^N]$ the role of the relevant configuration space for the abelian gauge theory is taken by the affine space

$$\Cal A_{B}^{(P,Q)} :=\{ A\in\Cal A^{(P)}\vert\quad \hat i_{\partial M}^{\ast}A=\phi
^{-1\ast}B\}\tag5.1.2$$ of those gauge potentials in $\Cal A^{(P)}$ whose restrictions to $\partial P$ equals the fixed connection $B$ under the
bundle isomorphism $\phi$. The tangent bundle of the configuration space is $T\Cal A_B^{(P,Q)}\cong\Cal A^{(P)}\times\Omega ^1(M,\partial
M;\frak t^N)$.\par

Let us consider the subgroup $\Cal G^{(M,\partial M)}=\{g\in\Cal G^{(P)}\vert\quad i_{\partial M}^{\ast}g=1\}$ of gauge transformations
approaching the unity on the boundary. This group gives a free action on $\Cal A_B^{(P,Q)}$ and correspondingly induces a smooth gauge orbit
space $\Cal M_B^{(P,Q)}=\Cal A_B^{(P,Q)}/\Cal G^{(M,\partial M)}$. In the following we will analyze the structure of the restricted gauge group
and exhibit the bundle structure of the space of connections and of the corresponding gauge orbit space. However, in the case $N=1$ some results
regarding the geometry of the gauge group have been already presented in [32].
\par
In order to characterize $\frak P_Q[M;\Bbb T^N]$ we consider the following long exact sequence in relative cohomology,

$$\multline\cdots @>>> H^1(\partial M;\Bbb Z^N) @>\hat\delta _1>>
H^2(M,\partial M;\Bbb Z^N) @>>>H^2(M;\Bbb Z^N)@>i_{\partial M}^{\ast}>> \\ @>i_{\partial M}^{\ast}>> H^2(\partial M;\Bbb Z^N) @>\hat\delta _2
>> H^3(M,\partial M;\Bbb Z^N)@>>>\cdots,\endmultline\tag5.1.3$$
where $\hat\delta _{\ast}$ are the connecting homomorphisms. Any two bundles $P_1,P_2\in\frak P_Q[M;\Bbb T^N]$ are related to each other by a
principal $\Bbb T^N$-bundle over $M$ whose Chern-class belongs to $H^2(M,\partial M;\Bbb Z^N)$. On the other hand the principal bundle $Q$ over
$\partial M$ can be extended to an principal bundle $P$ in $\frak P_Q[M;\Bbb T^N]$ if and only if $\delta _2 (Q)=0$. Generally the obstruction
belongs to $H^3(M,\partial M;\Bbb Z^N)$. If $Q$ is chosen to be the trivial bundle, then there is an isomorphism $\frak P_{Q=0}[M;\Bbb T^N]\cong
H^2(M,\partial M;\Bbb Z^N)$.\par

In order to proceed we need a brief digression on some results of Hodge theory on manifolds with a boundary [33-35]. Let us define the following
vector spaces of normal and tangential forms on $M$.

$$\align\Omega ^k(M,\partial M;\Bbb R)&=\Omega _{nor}^k(M;\Bbb R)
=\{ \alpha\in\Omega ^k(M,\Bbb R) \vert\quad i_{\partial M}^{\ast}\alpha =0\} \\ \Omega _{tan}^k(M,\Bbb R) & =\{\alpha\in\Omega ^k(M,\Bbb
R)\vert\quad i_{\partial M}^{\ast}\star\alpha =0\} \\ \Omega _{abs}^k(M;\Bbb R) &=\{\alpha\in\Omega ^k(M,\Bbb R)\vert\quad i_{\partial
M}^{\ast}\star\alpha =0\quad i_{\partial M}^{\ast}\star d\alpha =0\} \\ \Omega _{rel}^k(M;\Bbb R) &=\{\alpha\in\Omega ^k(M,\Bbb R)\vert\quad
i_{\partial M}^{\ast}\alpha =0\quad i_{\partial M}^{\ast}d^{\ast}\alpha =0\}.\tag5.1.4\endalign$$ For a compact, connected and oriented manifold
$M$ with boundary $\partial M$ one obtains

$$\split & <d\alpha ,\beta >-<\alpha ,d^{\ast}\beta > =\int\limits _{\partial
M}i_{\partial M}^{\ast}(\alpha\wedge\star\bar\beta )\\
& <\Delta\alpha ,\beta >-<d\alpha ,d\beta
>-<d^{\ast}\alpha ,d^{\ast}\beta > =\int\limits _{\partial
M}i_{\partial M}^{\ast}(d^{\ast}\alpha\wedge\star\beta -\beta\wedge\star d\alpha )\\ & <\Delta\alpha ,\beta >-<\alpha ,\Delta\beta >=\int
_{\partial M}i_{\partial M}^{\ast}(d^{\ast}\alpha\wedge\star\beta -d^{\ast}\beta\wedge\star\alpha +\alpha\wedge\star d\beta -\beta\wedge\star
d\alpha ).\endsplit\tag5.1.5$$ Let us define differential operators subjected to the different boundary conditions in (5.1.4), namely
$d_{k,nor}=d_k|_{\Omega ^k(M,\partial M;\Bbb R)}$ and $d_{k,tan}^{\ast}=(-1)^{k(n+1)+1}\star d_{n-k,nor}\star\colon\Omega _{tan}^k(M,\Bbb
R)\rightarrow\Omega _{tan}^{k-1}(M,\Bbb R)$. Accordingly, two different Laplacian operators can be distinguished, namely

$$\split\Delta
_{k}^{abs} &=d_{k-1}d_{k,tan}^{\ast}+d_{k+1,tan}^{\ast}d_{k}\colon\Omega _{abs}^{k}(M;\Bbb R) @>>>\Omega ^{k}(M;\Bbb R) \\ \Delta _k^{rel}
&=d_{k-1,nor}d_{k}^{\ast}+d_{k+1}^{\ast}d_{k,nor}\colon\Omega _{rel}^{k}(M;\Bbb R) @>>>\Omega ^{k}(M;\Bbb R),\endsplit\tag5.1.6$$ which are
elliptic and self adjoint on their respective domains of definition. On manifolds with a boundary there exist the following three kinds of Hodge
decompositions
$$\split \Omega ^k(M;\Bbb R)& = d\Omega _{nor}^{k-1}(M;\Bbb
R)\oplus d^{\ast}\Omega _{tan}^{k+1}(M;\Bbb R)\oplus\Cal H^k(M)\\
\Omega ^k(M;\Bbb R)& = d\Omega ^{k-1}(M;\Bbb R)\oplus
d^{\ast}\Omega _{tan}^{k+1}(M;\Bbb R)\oplus Harm_{abs}^k(M;\Bbb R)\\
\Omega ^k(M;\Bbb R)& = d\Omega _{nor}^{k-1}(M;\Bbb R)\oplus d^{\ast}\Omega ^{k+1}(M;\Bbb R)\oplus Harm_{rel}^k(M;\Bbb R),\endsplit\tag5.1.7$$
where $\Cal H^k(M)=\{\varpi\in\Omega ^k(M;\Bbb R)\vert\quad d\varpi =d^{\ast}\varpi =0\}$ is called the space of harmonic $k$-form fields. The
cohomology can be equivalently characterized by the kernel of the Laplace operators
$$\split Harm_{abs}^k(M;\Bbb R) &:=\ker\Delta _{k}^{abs}\cong
H^k(M;\Bbb R)\\ Harm_{rel}^k(M;\Bbb R) &:=\ker\Delta _{k}^{rel}\cong H^k(M,\partial M;\Bbb R).\endsplit\tag5.1.8$$ Furthermore one can define
the relative Green\rq s operator [32,33]

$$\gather G_{k}^{rel}\colon \Omega ^k(M;\Bbb R)
@>>>Harm_{rel}^{k}(M)^{\perp}\cap\Omega _{rel}^{k}(M;\Bbb R), \\
G_{k}^{rel}=( \Delta_{k}^{rel}|_{Harm_{rel}^{k}(M)^{\perp}})^{-1}\cdot\Pi ^{Harm_{rel}^{k}(M)^{\perp}},\tag5.1.9\endgather$$ satisfying $\Delta
_{k}^{rel}G_{k}^{rel} =\Pi ^{Harm_{rel}^{k}(M)^{\perp}}$, where $\Pi ^{Harm_{rel}^{k}(M)^{\perp}}$ is the projector onto the orthogonal
complement of $Harm_{rel}^{k}(M)$. The relative Green\rq s operator $G_{k}^{rel}$ commutes with both the differential $d$ and the
co-differential $d^{\ast}$. Analogously, it is possible to define the Green\rq s operator $G_{k}^{abs}$ for absolute boundary conditions.

\proclaim{Proposition 5.1} There exists an isomorphism between the abelian groups

$$\Cal G^{(M,\partial M)}\cong\Omega _{\Bbb
Z}^1(M,\partial M;\Bbb R^N).\tag5.1.10$$\endproclaim \demo{Proof} Let $C_{\ast}(M;\Bbb Z)$ ($C_{\ast}(M,\partial M;\Bbb Z)$) denote the complex
of smooth (relative) singular chains and let $Z_{\ast}(M,\partial M;\Bbb Z)$ be the subcomplex of relative cycles. Let $x_0\in\partial M$ be a
fixed point at the boundary. Given any $x\in M$ we choose a path $c_{x_0,x}$ in $M$ connecting $x_0$ with $x$. This path can be viewed as
1-chain in $C_1(M;\Bbb Z)$. Then the isomorphism $\kappa _{(M,\partial M)}\colon\Cal G^{(M,\partial M)}\rightarrow\Omega _{\Bbb Z}^1(M,\partial
M;\Bbb R^N)$ is provided by

$$\kappa _{(M,\partial M)}
(g)=\frac{1}{2\pi \sqrt{-1}}\ g^{\ast}\vartheta ,\qquad \kappa _{(M,\partial M)}^{-1}(\alpha )(x)=\exp{(2\pi\sqrt{-1}\int\limits
_{c_{x_0,x}}\alpha )}.\tag5.1.11$$

In order to prove that (5.1.11) is indeed well-defined, one chooses a different base point $x_0^{\prime}$ and a corresponding path
$c_{x_0^{\prime},x}$ connecting $x_0^{\prime}$ and $x$. Then the combined path $c_{x_0^{\prime},x}\diamond c_{x_0,x}^{-1}\in Z_1(M,\partial
M;\Bbb Z)$, since $\partial (c_{x_0^{\prime},x}\diamond c_{x_0,x}^{-1})\in Z_0(\partial M;\Bbb Z)$. The integral of any differential 1-form in
$\Omega _{\Bbb Z}^1(M,\partial M;\Bbb R^N)$ over this 1-cycle gives an integer. Analogously, the value of $\kappa _{(M,\partial M)}^{-1}$ is
independent of the actual path connecting $x_0$ and $x$. \qed
\enddemo

Like in the case of empty boundary, the gauge group $\Cal G^{(M,\partial M)}$ is not connected, which is displayed in the next statement.

\proclaim{Proposition 5.2} Let $\frak G^{(M,\partial M)}$ denote the Lie algebra of $\Cal G^{(M,\partial M)}$. Then the following sequence of
abelian groups is split exact:

$$0\rightarrow\frak G^{(M,\partial M)}@>\exp >>\Cal G^{(M,\partial
M)}@>\hat\kappa _{(M,\partial M)}^{\prime}>>H_{\Bbb Z}^1(M,\partial M;\Bbb R^N)\rightarrow 0,\tag5.1.12$$ where $\hat\kappa _{(M,\partial
M)}^{\prime} (g)=[\kappa _{(M,\partial M)} (g)]$.

\endproclaim
\demo{Proof} It is easy to show that (5.1.12) is exact. Let $\Pi ^{Harm_{rel}^{1}(M)}$ be the projector onto $Harm_{rel}^{1}(M)$ then the split
of the exact sequence is provided by the following isomorphism of abelian groups

$$\split & \hat\kappa _{(M,\partial M)}\colon H_{\Bbb Z}^1(M,\partial M;\Bbb
R^N)\times\frak G^{(M,\partial M)}\rightarrow\Cal G^{(M,\partial M)} \\ & \hat\kappa _{(M,\partial M)}([\alpha ],\xi )(x)=\exp{\xi
(x)}\cdot\exp{(2\pi\sqrt{-1}\int _{c_{x_0,x}}\Pi ^{Harm_{rel}^{1}(M)}(\alpha ))},\endsplit\tag5.1.13$$ where $c_{x_0,x}$ denotes a path in $M$
connecting $x_0$ and $x$. The independence of (5.1.13) of the selected path can be proved analogously than in the proof of proposition 5.1.
Hence any $g\in\Cal G^{(M,\partial M)}$ admits the following realization

$$g(x)=\exp{(G_0^{rel}d_1^{\ast}g^{\ast}\vartheta )(x)}\cdot\exp{(2\pi\sqrt{-1}\int _{c_{x_0,x}}
\Pi ^{Harm_{rel}^{1}(M)}(\frac{1}{2\pi\sqrt{-1}}\ g^{\ast}\vartheta ))}.\tag5.1.14$$ From (5.1.12) one finally obtains $\pi _0(\Cal
G^{(M,\partial M)})\cong H_{\Bbb Z}^1(M,\partial M;\Bbb R^N)$.\qed
\enddemo

The Lefschetz duality [31] states that the following isomorphisms exist

$$D_k^{(M,\partial M)}\colon H^k(M;\Bbb Z)\rightarrow H_{n-k}(M,\partial M;\Bbb
Z),\qquad\hat D_k^{(M,\partial M)}\colon H^k(M,\partial M;\Bbb Z) \rightarrow H_{n-k}(M;\Bbb Z),\tag5.1.15$$ implying $b_k^{rel}=b_{n-k}^{abs}$,
where $b_k^{rel}=\dim H^k(M,\partial M;\Bbb R)$ and $b_k^{abs}=\dim H^k(M,;\Bbb R)$. We choose a set of relative 1-cycles $\gamma _i^{rel}\in
Z_1(M,\partial M;\Bbb Z)$, for $i=1,\ldots ,b_1^{rel}$, whose homology classes $[\gamma _i^{rel}]$ provides a Betti basis for $H_1(M,\partial
M;\Bbb Z)$. Based on the following isomorphisms $\forall k=1,\ldots ,n$, namely

$$\split & H^k(M,\partial M;\Bbb Z)/TorH^k(M,\partial M;\Bbb Z)
\cong H_{\Bbb Z}^k(M,\partial M;\Bbb R)
\cong Harm_{\Bbb Z}^k(M,\partial M;\Bbb R)\\
& H^k(M;\Bbb Z)/TorH^k(M;\Bbb Z)\cong H_{\Bbb Z}^k(M;\Bbb R) \cong Harm_{abs,\Bbb Z}^k(M;\Bbb R)\endsplit\tag5.1.16$$ a basis of harmonic forms
$(\varrho _i^{(abs,n-1)})_{i=1}^{b_{n-1}^{abs}}\in Harm_{abs,\Bbb Z}^{n-1}(M;\Bbb R)$ can be selected from the cohomology basis
$D_1^{-1}([\gamma _i^{rel}])$ of $H^{n-1}(M;\Bbb Z)$. The product

$$\split H^1(M;\Bbb Z)/TorH^1(M;\Bbb Z) &\times
H^{n-1}(M,\partial M;\Bbb Z)/TorH^{n-1}(M,\partial M;\Bbb Z) \rightarrow\Bbb Z\\  (\mu ,\nu )&\mapsto <\mu , D_{n-1}^{(M,\partial M)}(\nu
)>=<\mu\cup\nu ,[M]>,\endsplit\tag5.1.17$$ gives a perfect pairing like in the boundary-less case. Thus a basis $(\varrho
_i^{(rel,1)})_{i=1}^{b_1^{rel}}\in Harm_{\Bbb Z}^1(M,\partial M;\Bbb R)$ can be adjusted in such a way so that

$$\int _{\gamma _j^{rel}}\varrho
_i^{(rel,1)} =\int _M\ \varrho _i^{(rel,1)}\wedge\varrho _j^{(abs,n-1)}=\delta _{ij}.\tag5.1.18$$ This basis induces a metric
$$h_{jk}^{rel}=<\varrho _j^{(rel,1)},\varrho _k^{(rel,1)}>\tag5.1.19$$
on $Harm_{rel}^1(M;\Bbb R^N)$. Moreover for any $[\alpha ]\in H^1(M,\partial M;\Bbb R)$ the following relation holds $\int _{\gamma
_j^{rel}}\alpha =\int _M\alpha\wedge\varrho _j^{(abs,n-1)}$.

Let us now choose an arbitrary but fixed background gauge field $A_0\in\Cal A_B^{(P,Q)}$ and define the smooth surjective map $\pi _{\Cal
M_B^{(P,Q)}}^{A_0}\colon\Cal M_B^{(P,Q)}\rightarrow\Bbb T^{b_1N^{rel}}$ by

$$\pi _{\Cal M_B^{(P,Q)}}^{A_0}([A])=(e^{\int _M (A-A_0)\wedge\varrho
_1^{(abs,n-1)}},\ldots , e^{\int _M (A-A_0)\wedge\varrho _{b_1^{rel}}^{(abs,n-1)}}).\tag5.1.20$$ In terms of the inner product (3.1) the
components in (5.1.20) can be rewritten in the form

$$\int _M (A-A_0)\wedge\varrho _j^{(abs,n-1)}=(-1)^n<A-A_0,\star
\varrho _j^{(abs,n-1)}>.\tag5.1.21$$ Let us remark that $\star\varrho _j^{(abs,n-1)}\in Harm_{rel}^1(M;\Bbb R)$ since the Hodge operator
provides an isomorphism between the relative and absolute harmonic forms.\par

By (5.1.20) we are able to construct a finite open cover $\tilde{\Cal U}=\{\tilde{U}_a\}$ of the infinite dimensional manifold $\Cal
M_{B}^{(P,Q)}$ by defining $\tilde{U}_a=(\pi_{\Cal M_B^{(P,Q)}}^{A_0})^{-1}(V_a)$, with $a=1,\ldots ,2^{b_1^{rel}N}$. Then the geometrical
structure of the bundle of gauge potentials is displayed by the following:

\proclaim{Theorem 5.3} $\Cal A_B^{(P,Q)}$ is a flat principal bundle over $\Cal M_B^{(P,Q)}$ with structure group $\Cal G^{(M,\partial M)}$ and
projection $\pi _{\Cal A_B^{(P,Q)}}$. This bundle is trivializable if $H^1(M,\partial M,\Bbb Z)\cong H_{n-1}(M;\Bbb Z)=0$.
\endproclaim

\demo{Proof} A bundle atlas is provided by the following family of local trivializations $\tilde\varphi _{a}^{A_0}\colon \tilde U_{a}\times\Cal
G^{(M,\partial M)}\rightarrow\pi _{\Cal A_B^{(P,Q)}}^{-1}(\tilde U_a)$, $\tilde\varphi _{a}^{A_0}([A],g)=A^{\tilde\omega _{a}^{A_0}(A)^{-1}g}$,
where

$$\split &\tilde\omega _{a}^{A_0}\colon\pi _{\Cal A_B^{(P,Q)}}^{-1}(\tilde U_a)
\rightarrow \Cal G^{(M,\partial M)}\\ &\tilde\omega _{a}(A) = \hat\kappa _{(M,\partial M)} ([\sum _{j=1}^{b_1^{rel}}\tilde\epsilon
_{a_{j}}^{A_0}(A)\varrho
_j^{(rel,1)}],\exp{G_0 ^{rel}d_1^{\ast}(A-A_0)}),\\
&\tilde\epsilon _{a_{j}}^{A_0}(A)=(\tilde{\epsilon} _{a_{j1}}^{A_0}(A),\ldots ,\tilde{\epsilon} _{a_{j\alpha }}^{A_0}(A),\ldots
,\tilde{\epsilon} _{a_{jN}}^{A_0}(A)) \colon\pi
_{\Cal A_B^{(P,Q)}}^{-1}(\tilde U_{a})\rightarrow\Bbb Z^{N}\\
&\tilde{\epsilon} _{a_{j\alpha }}^{A_0}(A)=\frac{1}{2\pi\sqrt{-1}}\int _M ((A^{\alpha}-A_0^{\alpha})\wedge\varrho _j^{(abs,n-1)})-
s_{a_{j\alpha}}(e^{\int _M (A^{\alpha}-A_0^{\alpha})\wedge\varrho _j^{(abs,n-1)}})\endsplit\tag5.1.22$$ Since $\frac{1}{2\pi\sqrt{-1}}\
g^{\ast}\vartheta$ has integer periods its projection onto the space of relative harmonic one-forms with integer periods is given by

$$\Pi
^{Harm_{rel}^1(M)}(\frac{1}{2\pi\sqrt{-1}}\ g^{\ast}\vartheta )= \sum _{j=1}^{b_1^{rel}}(h_{jk}^{rel})^{-1}<\frac{1}{2\pi\sqrt{-1}}\
g^{\ast}\vartheta ,\varrho _j^{(rel,1)}>\varrho _k^{(rel,1)}=\sum _{j=1}^{b_1^{rel}}m_j\varrho _j^{(rel,1)},\tag5.1.23$$ where $m_k\in\Bbb Z^N$.
Using that $\tilde\epsilon _{a_k}^{A_0}(A^g)=\tilde\epsilon _{a_k}^{A_0}(A)+m_k$ and (5.1.15) one derives $\tilde\omega _{a}^{A_0}
(A^g)=\tilde\omega _a^{A_0}(A)g$. The transition functions $\tilde\varphi _{aa^{\prime}}^{A_0}\colon \tilde U_a\cap \tilde
U_{a^{\prime}}\rightarrow\Cal G^{(M,\partial M)}$ yield

$$\tilde\varphi _{aa^{\prime}}^{A_0}([A])=\hat\kappa _{(M,\partial M)}([\sum
_{j=1}^{b_1^{rel}} g_{a_ja_j^{\prime}}^{\Bbb T^N}(e^{\int _M (A-A_0)\wedge\varrho _j^{(abs,n-1)}})\varrho _j^{(rel,1)}],0)\tag5.1.24$$ and are
locally constant. Together with the universal coefficient theorem one concludes that the bundle is trivializable if $H^1(M,\partial M;\Bbb
Z)=0$. Moreover the transition functions are locally constant. Like in the proof of theorem 4.3 on can easily verify that a different choice for
the background gauge field $A_0$ would lead to an equivalent bundle atlas. \qed\enddemo

\proclaim{Theorem 5.4} For each arbitrary but fixed connection $A_0\in\Cal A_B^{(P,Q)}$, the manifold $\Cal M_B^{(P,Q)}$ admits the structure of
a trivializable vector bundle over $\Bbb T^{b_1^{rel}N}$ with projection $\pi _{\Cal M_B^{(P,Q)}}^{A_0}$ and typical fibre $\Cal N^{(M,\partial
M)}:=(imd_2^{\ast}\cap\Omega _{rel} ^1(M;\Bbb R))\otimes\frak t^N$.
\endproclaim

\demo{Proof} A bundle atlas is provided by the local diffeomorphism

$$\align &\tilde\chi _a^{A_0} \colon V_a\times\Cal N^{(M,\partial M)}
\rightarrow\Cal M_B^{(P,Q)}\\ &\chi _{a}^{A_0}(\vec z_1,\ldots ,\vec z_{b_1^{rel}},\tau ) =[A_0+2\pi\sqrt{-1}\sum
_{j=1}^{b_1^{rel}} s_{a_j}(\vec z_j)\varrho _j^{(rel,1)}+\tau ]\\
&(\tilde\chi _a^{A_0})^{-1}([A]) =(\pi _{\Cal M_B^{(P,Q)}}^{A_0}([A]), d_2^{\ast}G_2^{rel}(F_A-F_{A_0})).\tag5.1.25\endalign$$ There exists a
unique vector bundle structure induced by the bundle chart $\tilde\chi _a^{A_0}$. In analogy with theorem 4.4 we can easily prove that the
choice of a different background gauge field induces an isomorphic vector bundle structure on $\Cal M_B^{(P,Q)}$. \qed\enddemo

The topological structure of the true configuration space is characterized by the following:

\proclaim{Corollary 5.5} There exist the following isomorphisms
$$\split & H^k(\Cal M_B^{(P,Q)};\Bbb Z)\cong H^k(\Bbb
T^{b_1^{rel}N};\Bbb Z)=\Bbb Z^{\binom {b_1^{rel}N}k}\\ &\pi _k(\Cal M_B^{(P,Q)})\cong\pi _k(\Bbb T^{b_1^{rel}N})=\delta _{k1}\Bbb
Z^{b_1^{rel}N}.\endsplit\tag5.1.26$$\qed\endproclaim
\bigskip\bigskip
{\bf 5.2. The partition function, VEV of gauge invariant observables and the Green\rq s functions}\bigskip

Now we are ready to apply the previous results to the construction of the partition function, the VEV of gauge invariant observables, Green\rq s
functions and the field strength correlation functions in the case of manifolds with a boundary. Much of the results which have been elaborated
for closed manifolds can be directly generalized, if the boundary conditions are appropriately specified. Thus we are going to skip the details
and to present the results only.\par

We introduce a partition of unity $\{\tilde p_{a}\}$ for $\Cal M_{B}^{(P,Q)}$ subordinate to $\tilde\Cal U$. Thereby $\tilde p_{a}:=\pi _{\Cal
M_{B}^{(P,Q)}}^{\ast}\tilde p_{a}^{\prime}$, where $\tilde p_{a}^{\prime}:=\prod _{\alpha =1}^N\prod
_{j=1}^{b_1^{rel}}q_{j\alpha}^{\ast}\hat{p}_{a_{j\alpha}}$ with the multi-index $a=(a_{11},\ldots ,a_{b_1^{rel}N})\in\Bbb Z_2^{b_1N^{rel}}$.
\par
Let $\Phi _i=\Phi _i(A)$, $i=0,1,2$ denote three functionals of the gauge fields which will be specified later on. For each $\Phi _i$ we choose
an associated source $J_i\in\Omega _{rel} ^{\ast}(M;\Bbb R)\otimes\frak t^N$, $i=1,2$ and introduce a generating functional by

$$Z_{A_0}^{(P,Q)}[J;B,\Phi _i ]=\int _{\Cal A_{B}^{(P,Q)}}vol_{\Cal
A_{B}^{(P,Q)}}\sum _{a\in\Bbb Z_2^{b_1^{rel}N}}(\pi _{\Cal A_B^{(P,Q)}}^{\ast}\tilde p_a) \ e^{-S_{inv}-(\tilde\omega _a^{A_0})^{\ast}\tilde
S_{gf}+<\Phi _i ,J_i>}.\tag5.2.1$$ This gives rise to the following correlation functions

$$\Cal V_q^{(P,Q)}(v_1\ldots ,v_q;\Phi _i) :=
\frac{\partial ^q}{\partial t_1\cdots\partial t_q} \vert _{t_1=\ldots =t_q=0} \ \frac{Z_{A_0}^{(P,Q)}[\sum _{i=1}^qt_iv_i;B,\Phi
_i]}{Z_{A_0}^{(P,Q)}[0;B,\Phi _i]},\tag5.2.2$$ where $v_1,\ldots ,v_q\in \Omega _{rel} ^{\ast}(M;\Bbb R)\otimes\frak t^N$. The form degree
depends on which of the following three cases is considered: \roster \item $i=0$: If $\Phi _0(A)=0$, eq. (5.2.1) reduces to the partition
function of the theory, denoted by $Z_{A_0}^{(P,Q)}(B)$, \item $i=1$: If $\Phi _1(A)=A-A_0$ and $J_1\in\Omega _{rel} ^1(M;\Bbb R)\otimes\frak
t^N$, eq. (5.2.2) gives the generating functional for the $q$-th point Green\rq s functions, denoted by $\Cal S_q^{(P,Q)}$, \item $i=2$: If
$\Phi _2(A)=F_A$ and $J_2\in\Omega _{rel} ^2(M;\Bbb R)\otimes\frak t^N$, eq. (5.2.2) gives the VEV of the field strength polynomial of degree
$q$, denoted by $\Cal W_q^{(P,Q)}$.\endroster The next step is to find an appropriate choice for the regularizing function in (2.1): Let us
denote by $\theta ^{(M,\partial M)}$ the Maurer Cartan form on $\Cal G^{(M,\partial M)}$. The induced left-invariant volume form is given by
$vol_{\Cal G^{(M,\partial M)}}=\left( \det{(\bar\theta\theta)}^{1/2} \Cal Dg \right)$. A regularization of the volume of the gauge group $\Cal
G^{(M,\partial M)}$ will be provided by the gauge fixing function $\tilde S_{gf}$

$$e^{-\tilde S_{gf}(g)}=\frac{e^{-\tilde S_{gf}^{\prime}(g)}}{\int _{\Cal
G^{(M,\partial M)}} \ vol_{\Cal G^{(M,\partial M)}}\ e^{-\tilde S_{gf}^{\prime}}},\tag5.2.3$$ with an auxiliary gauge fixing function

$$\tilde S_{gf}^{\prime}(g)= \frac{1}{2}\Vert d^{\ast}g^{\ast}\vartheta\Vert _{\lambda}^2+
\frac{1}{2}\Vert \Pi ^{Harm_{rel}^1(M)}(g^{\ast}\vartheta )\Vert _{\lambda}^2.\tag5.2.4$$ Let $\Pi ^{imd_{0,nor}}:=d_{0,nor}G_0^{rel}d_1^{\ast}$
denote the projector onto the space of exact 1-forms on $M$ with normal boundary conditions and let $\tilde
K_{j}^{\alpha}(J_1)=-\sqrt{-1}<J_1^{\alpha},\sqrt{-1}\varrho _j^{(rel,1)}
>$ with $j=1,\ldots ,b_1^{rel}$ and $\alpha =1,\ldots ,N$ be regarded as
$b_1^{rel}N$-dimensional complex vector $\tilde K(J_1)$. Using the results of proposition 5.2 a straightforward calculation leads to:

\proclaim{Lemma 5.7} For the auxiliary gauge fixing function $\tilde S_{gf}^{\prime}$ (5.2.4) one gets

$$\multline \int _{\Cal G^{(M,\partial M)}} vol_{\Cal G^{(M,\partial M)}}
e^{-\tilde S_{gf}^{\prime}(g)+<J_1,g^{\ast}\vartheta >} =\\ = (\det{\Delta _0^{rel}|_{imd_1^{\ast}}})^{-N}\cdot\Theta _{b_1N^{rel}}(\tilde
K(J_1)|2\pi\sqrt{-1}\tilde\Lambda )e^{1/2<\Pi ^{imd_{0,nor}}(J_1),G_1^{rel}\Pi ^{imd_{0,nor}}(J_1)>},\endmultline\tag5.2.5$$ where
$\tilde\Lambda =\lambda\otimes h^{rel}$ is the tensor product of the matrix $(\lambda _{\alpha ,\beta} )_{\alpha ,\beta =1}^N$ of coupling
constants and the metric on the harmonic relative 1-forms $(h _{jk}^{rel} )_{j,k=1}^{b_1^{rel}}$.  \qed\endproclaim

Based on the choice of the gauge fixing function (5.2.3), the partition function can be displayed in the gauge field space as follows:

\proclaim{Proposition 5.8} Let $P\in\frak P_Q[M;\Bbb T^N]$ and let $B\in\Cal A^{(Q)}$ be an arbitrary but fixed connection. The partition
function for the abelian gauge theory with the classical action (3.6) on a manifold with a non-empty boundary is given by

$$Z_{A_0}^{(P,Q)}(B)=
\int _{\Cal A_B^{(P,Q)}} vol_{\Cal A_B^{(P,Q)}}\ \tilde\Cal F(A)\cdot
e^{-\frac{1}{2}(\|F_{A}\|_{\lambda}^2+\|d^{\ast}(A-A_0)\|_{\lambda}^2)},\tag5.2.6$$ with the positive definite functional

$$\multline\tilde\Cal F(A)=\\=(\det{\Delta _0^{rel}|_{imd_1^{\ast}}})^{N}
\Theta _{b_1^{rel}N}(0|2\pi\sqrt{-1}\Lambda ^{rel} )^{-1}\sum _{a\in\Bbb Z_2^{b_1^{rel}N}} (\pi _{\Cal A_B^{(P,Q)}}^{\ast}\tilde p_a) e^{-2\pi
^2\sum\limits _{\alpha ,\beta =1}^{N}\sum \limits _{j,k=1}^{b_1^{rel}} \lambda _{\alpha\beta}h_{jk}^{rel}\tilde\epsilon
_{a_{j\alpha}}^{A_0}(A)\tilde\epsilon _{a_{k\beta}}^{A_0}(A)},\endmultline\tag5.2.7$$ where the multi-index reads $a=(a_{11},\ldots
,a_{b_1^{rel}N})\in\Bbb Z_2^{b_1^{rel}N}$. \qed\endproclaim

With respect to the local bundle trivializations $\tilde\psi _a^{A_0}=\tilde \varphi _a^{A_0}\circ (\tilde{\chi _a^{A_0}}\times\Bbb I)$ the
volume form becomes

$$(\tilde\psi _a^{A_0})^{\ast}vol_{\Cal A_B^{(P,Q)}}=(\det h^{rel})^{N/2}\det{(\Delta
_0^{rel}|_{imd_1^{\ast}})}^{N/2} \ vol_{\Bbb T^{b_1N^{rel}}}\vert _{V_a}\wedge vol_{\Cal N^{(M,\partial M)}}\wedge vol_{\Cal G^{(M,\partial
M)}},\tag5.2.8$$ where $vol_{\Bbb T^{b_1^{rel}N}}=(\sqrt{-1})^{-b_1^{rel}N}q_{11}^{\ast}\vartheta ^{\Bbb T^1}\wedge\ldots \wedge
q_{b_1^{rel}N}^{\ast}\vartheta ^{\Bbb T^1}$ is the induced volume form on $\Bbb T^{b_1^{rel}N}$ restricted to a single patch $V_a$. Like in the
boundary-less case, the flat metric on $\Cal N^{(M,\partial M)}$ induces a volume form $vol_{\Cal N^{(M,\partial M)}}$ which formally is just
$\Cal D\tau $.\par

Given any background connection $A_0\in\Cal A_B^{(P,Q)}$, then the gauge field $A_0^{\prime}:=A_0-G_1^{rel}d_2^{\ast}F_{A_0}$ fulfills
$d_2^{\ast}F_{A_0^{\prime}}=0$. Hence we can restrict ourselves to the class of background gauge fields which satisfy the classical field
equation. Using (5.2.8) and that $\det{(\Delta _{0}^{rel}|_{imd_1^{\ast}})}=\det{(\Delta _{1}^{rel}|_{imd_{0,nor}})}$ a direct calculation
finally gives:

\proclaim{Proposition 5.9} Let $P\in\frak P_Q[M;\Bbb T^N]$ and let $B\in\Cal A^{(Q)}$ be an arbitrary but fixed connection. The generating
functional corresponding to the three cases is given by \roster\item $i=0$:
$$Z_{A_0}^{(P,Q)}(B)=(2\pi )^{b_1^{rel}N}(\det
h^{rel})^{N/2} (\det{\Delta _0^{rel}|_{imd_1^{\ast}}})^{N}(\det{\Delta _1^{rel}|_{Harm_{rel}^1(M)^{\perp}}})^{-N/2}
e^{-\frac{1}{2}\|F_{A_0}\|_{\lambda}^2},\tag5.2.9$$\item $i=1$:

$$\split Z_{A_0}^{(P,Q)}[J_1;B,&\Phi _1]
=(\det h^{rel})^{N/2}(\det{\Delta_0^{rel}|_{imd_1^{\ast}}})^{N}(\det{\Delta_1^{rel}|_{Harm_{rel}^1(M)^{\perp}}}) ^{-N/2}\frac{\Theta
_{b_1^{rel}N}(\tilde K(J_1)|2\pi\sqrt{-1}\tilde\Lambda )}{\Theta _{b_1^{rel}N}(0|2\pi\sqrt{-1}\tilde\Lambda )} \\ & \times
e^{-\frac{1}{2}\|F_{A_0}\|_{\lambda }^2}\cdot e^{\frac{1}{2}<J_1,G_1^{rel}J_1>_{\lambda ^{-1}}}\int\limits _{\Bbb T^{b_1N^{rel}}}vol_{\Bbb
T^{b_1^{rel}N}}\sum _{a_{11}=1}^2\ldots\sum _{a_{b_1^{rel}N}=1}^2 q_{11}^{\ast}\hat{p}_{a_{11}}\cdots
q_{b_1^{rel}N}^{\ast}\hat{p}_{a_{b_1^{rel}N}}\\
&\times e^{2\pi\sum _{\alpha =1}^N\sum _{j=1}^{b_1^{rel}}q_{j\alpha}^{\ast}s_{a_{j\alpha}}<J_1^{\alpha},\sqrt{-1}\rho
_j^{(rel,1)}>},\endsplit\tag5.2.10$$ \item $i=2$:
$$\split\tilde Z_{A_0}^{(P,Q)}[J_2;B,\Phi _2]= &(2\pi )^{b_1^{rel}N}(\det h^{rel})^{N/2}
(\det{\Delta _0|_{imd_1^{\ast}}})^{N}(\det{\Delta
_1|_{Harm_{rel}^{1}(M)^{\perp} }})^{-N/2}\\
&\times e^{-\frac{1}{2}\Vert F_{A_0}\Vert _{\lambda}^2+\frac{1}{2}<J_2,(\Pi ^{Harm_{rel}^2(M)^{\perp}}-d_{3}^{\ast}G_3^{rel}d_2)J_2>_{\lambda
^{-1}}+<F_{A_0},J_2>}.\endsplit\tag5.2.11$$
\endroster\qed\endproclaim

\proclaim{Proposition 5.10} Let $P\in\frak P_Q[M;\Bbb T^N]$ and let $B\in\Cal A^{(Q)}$ be an arbitrary but fixed connection. The VEV of any
gauge invariant function $f$ takes the form

$$\multline <f>_{(P,Q)}=\\=(\det{\Delta _0^{rel}|_{imd_1^{\ast}}})^{-N/2}\cdot
(\det{\Delta _1^{rel}|_{Harm_{rel}^1(M)^{\perp}}})^{N/2}\int\limits _{\Cal N^{(M,\partial M)}}\ \Cal D\tau\hat{f}_{(0,\ldots ,0)}(A_0+\tau
)\cdot e^{-\frac{1}{2}<\tau ,\Delta _1^{rel}|_{imd_1^{\ast}}\tau
>_{\lambda}},\endmultline\tag5.2.12$$ with the Fourier components
$$\hat{f}_{(0,\ldots ,0)}(A_0+\tau )=
\int\limits _{0}^{1}\cdots\int\limits _{0}^{1}dt_{11}\ldots dt_{b_1^{rel}N}\hat{f}(e^{2\pi\sqrt{-1}t_{11}},\ldots
,e^{2\pi\sqrt{-1}t_{b_1^{rel}N}},A_0+\tau ).\tag5.2.13$$ \endproclaim \demo{Proof} Eq. (5.2.12) follows from (2.4) by a direct calculation where
we follow the lines of section 4.2. Here any gauge invariant function $f$ on $\Cal A_{B}^{(P,Q)}$ projects to a global function
$\hat{f}=(\tilde\chi _a^{A_0})^{\ast}f$ on $\Bbb T^{b_1^{rel}N}\times\Cal N^{(M,\partial M)}$, which admits a Fourier series expansion.
\qed\enddemo

Let us define the field independent factor
$$\split\tilde\varepsilon _{j\alpha
,k\beta}^{(2)} &=\int _{\Bbb T^{b_1^{rel}N}} vol_{\Bbb T^{b_1^{rel}N}} \sum _{a_{11}=1}^2\ldots\sum _{a_{b_1^{rel}N}=1}^2 q_{11}^{\ast}\hat
p_{a_{11}}\cdots q_{b_1^{rel}N}^{\ast}\hat p_{a_{b_1^{rel}N}} s_{a_{j\alpha}}s_{a_{k\beta}}= \\ &= \cases (2\pi
)^{b_1^{rel}N-1}(\frac{17\pi}{12}-\frac{1}{\pi}), &\text{for $j=k$ and $\alpha =\beta$}\\ (2\pi )^{b_1^{rel}N-2}(\frac{3\pi}{2})^2, &\text{for
$j\neq k$ or $\alpha\neq\beta$ or both},\endcases
\endsplit\tag5.2.14$$ then a lengthy calculation gives:

\proclaim{Proposition 5.11} Let $P\in\frak P_Q[M;\Bbb T^N]$ and let $B\in\Cal A^{(Q)}$ be an arbitrary but fixed connection. The $q$-point
Green\rq s functions of the gauge fields read:\par 1) One-point function:

$$\Cal S_1^{(P,Q)}(v ) =\sum _{\alpha =1}^N\sum
_{j=1}^{b_1^{rel}}\varepsilon _{j\alpha}^{(1)}<v^{\alpha},\sqrt{-1}\varrho _j^{(rel,1)}>+ \frac{d}{dt}|_{t=0}\ln{\Theta _{b_1^{rel}N}(\tilde
K(t\eta )|2\pi\sqrt{-1}\tilde\Lambda )}.\tag5.2.15$$\par 2) Two-point function:

$$\split &\Cal S_2^{(P,Q)}(v_1,v_2) = <v_1,
G_1^{rel}v_2>_{\lambda ^{-1}} \\ &+\sum _{\alpha =1}^N\sum _{j=1}^{b_1^{rel}}\varepsilon _{j\alpha}^{(1)}<v_{1}^{\alpha},\sqrt{-1}\varrho
_j^{(rel,1)}>(\frac{d}{dt}|_{t=0}\ln{\Theta _{b_1^{rel}N}(\tilde
K(tv_2 )|2\pi\sqrt{-1}\tilde\Lambda )}) \\
&+ \sum _{\alpha =1}^N\sum _{j=1}^{b_1^{rel}}\varepsilon _{j\alpha}^{(1)}<v _{2}^{\alpha},\sqrt{-1}\varrho
_j^{(rel,1)}>(\frac{d}{dt}|_{t=0}\ln{\Theta _{b_1^{rel}N}(\tilde K(tv_1 )|2\pi\sqrt{-1}\tilde\Lambda )})\\ &+ (2\pi )^{2-b_1^{rel}N}\sum
_{\alpha ,\beta =1}^N\sum _{j,k=1}^{b_1^{rel}}\tilde\varepsilon _{j\alpha ,k\beta}^{(2)}<v_{1}^{\alpha},\sqrt{-1}\varrho _j^{(rel,1)}><v
_{2}^{\beta},\sqrt{-1}\varrho _k^{(rel,1)}>\\
&+\Theta _{b_1^{rel}N}(0|2\pi\sqrt{-1}\tilde\Lambda )^{-1}\frac{\partial ^2}{\partial t_1\partial t_2} \vert _{t_1=t_2=0}\Theta
_{b_1N^{rel}}(\tilde K(\sum _{l=1}^2t_lv_l)|2\pi\sqrt{-1}\tilde\Lambda ).\endsplit\tag5.2.16$$\qed\endproclaim

We remark that the Green\rq s functions are already independent of the fixed component $\Cal A_B^{(P,Q)}$. Let us define the following
abbreviations
$$\align \mu_{ij}^{rel} &=<v_i,(\Pi
^{Harm_{rel}^2(M)^{\perp}}-d_{3}^{\ast}G_3^{rel}d_2)v_j>_{\lambda ^{-1}}\\
\nu _i^{rel} &= <F_{A_0},v_i>,\tag5.2.18\endalign$$ for $v_i,v_j\in\Omega ^2(M;\frak t^N)$ and let $\varsigma$ denote a permutation of the
indices $\{1,\ldots, q\}$ then a lengthy calculation gives the following:

\proclaim{Proposition 5.12} Let $P\in\frak P_Q[M;\Bbb T^N]$ and let $B\in\Cal A^{(Q)}$ be an arbitrary but fixed connection. The correlation
functions $\Cal W_q^{(P,Q)}$ of the field strength are given by
\par 1) $q=2k$

$$\split \Cal W_{2k}^{(P,Q)} (v_1\ldots ,&v_{2k})=\frac{1}{(2k)!}
\sum _{\varsigma} \nu _{\varsigma (1)}^{rel}\cdots\nu _{\varsigma (2k)}^{rel}\\
& +\sum _{l=1}^{k-1}\frac{1}{2^ll!(2(k-l))!} \sum _{\varsigma} \mu _{\varsigma (1)\varsigma (2)}^{rel}\cdots\mu _{\varsigma (2l-1)\varsigma
(2l)}^{rel}\nu _{\varsigma (2l+1)}^{rel}\cdots\nu _{\varsigma (2k)}^{rel}\\
&+\frac{1}{2^kk!}\sum _{\varsigma}\mu _{\varsigma (1)\varsigma (2)}^{rel}\cdots\mu _{\varsigma (2k-1)\varsigma (2k)}^{rel}
\endsplit\tag5.2.18$$

2) For $q=2k+1$

$$\split \Cal W_{2k+1}^{(P,Q)} (v_1 &\ldots ,v_{2k+1})=\frac{1}{(2k+1)!}
\sum _{\varsigma}\nu _{\varsigma (1)}^{rel}\cdots\nu _{\varsigma (2k+1)}^{rel}\\
 &+\sum _{l=1}^{k}\frac{1}{2^ll!(2(k-l)+1)!}\sum _{\varsigma} \mu
_{\varsigma (1)\varsigma (2)}^{rel}\cdots\mu _{\varsigma (2l-1)\varsigma (2l)}^{rel}\nu _{\varsigma (2l+1)}^{rel}\cdots\nu _{\varsigma
(2k+1)}^{rel},\endsplit\tag5.2.19$$ where $v_1,\ldots v_q\in\Omega ^2(M;\frak t^N)$.\qed
\endproclaim

For $H^1(M,\partial M;\Bbb Z)=0$ and the trivial $\Bbb T^N$-bundle $M\times \Bbb T^N$ over $M$ one can choose $A_0=0$. For consistency, $Q$ must
be also trivializable with trivial connection $B=0$. In that case the bundle structure reduces to $\Cal M_B^{(P,Q)}\cong \Cal N^{(M,\partial
M)}$ implying that the orbit space consists of transversal fields with relative boundary conditions only. Under these assumptions the partition
function reduces to

$$Z_{A_0=0}^{(P,Q)}[B=0]=(\det{\Delta _0^{rel}})^{N}(\det{\Delta
_1^{rel}})^{-N/2},\tag5.2.20$$ which for $N=1$ gives the covariant expression for the functional integral of quantized Maxwell theory (e.g. see
[36]).\par

\subheading{The partition function on $\Cal A_B^{(M,Q)}$} The set of all connections on $M$, denoted by $\Cal A_B^{(M,Q)}$, is the disjoint
union

$$\Cal A_B^{(M,Q)}=\bigsqcup _{P\in\Cal P_Q[M;\Bbb T^N]}\Cal
A_B^{(P,Q)}.\tag5.2.21$$ Let us now discuss the special case if $Q$ is the trivial principal bundle over $\partial M$ with trivial flat
connection $B=0$. Any $P\in\frak P_{Q=0}[M;\Bbb T^N]$ is uniquely characterized by a Chern class $c=(c^{\alpha})_{\alpha =1}^{N}\in
H^2(M,\partial M;\Bbb Z^N)$ which can be written in the form

$$c^{\alpha}=\sum _{j=1}^{b_2^{rel}}m_{j\alpha}c_{j}^{(rel,2)}+
\sum _{k=1}^{\tilde r}t_{k\alpha}y_{k}^{(rel,2)},\tag5.2.22$$ where $(c_{j}^{(rel,2)})_{j=1}^{b_2^{rel}}$ denotes a Betti basis of
$H^2(M,\partial M;\Bbb Z)$ with rank $b_2^{rel}$ and $m_{j\alpha}\in\Bbb Z$ for $j=1,\ldots ,b_2^{rel}$, $\alpha =1,\ldots , N$. Furthermore
$(y_{k}^{(rel,2)})_{k=1}^{\tilde r}$ is assumed to generate $TorH^2(M,\partial M;\Bbb Z)$ with torsion coefficients $\tilde l_k$, i.e. $\tilde
l_ky_{k}^{(rel,2)}=0$ and $t_{k\alpha}\in\Bbb Z_{\tilde l_k}$ for $k=1,\ldots ,\tilde r$, $\alpha =1,\ldots ,N$. Accordingly $F_{A_0}\in
Harm_{rel,\Bbb Z}^2(M)\otimes\frak t^N$, where $Harm_{rel,\Bbb Z}^{2}(M)$ denotes the space of harmonic relative differential 2-forms with
integer periods. Let $\varrho _j^{(rel,2)}$ ($j=1,\ldots ,b_2^{rel}$) be a basis of $Harm_{rel,\Bbb Z}^{2}(M)$ then

$$F_{A_0}^{\alpha}=2\pi\sqrt{-1}\sum _{j=1}^{b_2^{rel}}m_{j\alpha}\rho
_j^{(rel,2)},\quad m_j^{\alpha}\in\Bbb Z.\tag5.2.23$$ Let $h_{jk}^{(rel,2)}=<\varrho _j^{(rel,2)},\varrho _k^{(rel,2)}>$ be the induced metric
on the harmonic 2-forms then one finds:

\proclaim{Proposition 5.13} Let $Q$ be the trivial $\Bbb T^N$-bundle over $\partial M$. The partition function, denoted by $Z^{(M,Q)}[B=0]$, on
the total configuration space $\Cal A_B^{(M,Q)}$ is then given by

$$\split Z^{(M,Q)}[B=0] &=(2\pi )^{b_1^{rel}N}(\det{
h^{rel}})^{N/2} (\det{\Delta _0^{rel}|_{imd_1^{\ast}}})^{N}
(\det{\Delta _1|_{Harm_{rel}^1(M)^{\perp}}})^{-N/2}\\
&\times\Theta _{b_2^{rel}N}(0|2\pi\sqrt{-1}\Lambda ^{(rel, 2)})\cdot ord(TorH^2(M,\partial M;\Bbb Z^N)),\endsplit\tag5.2.24$$ where $\Lambda
^{(rel, 2)}=\lambda\otimes h_{jk}^{(rel,2)}$. \qed\endproclaim

The partition function $Z^{(M,Q)}[B=0]$ does not depend on the explicit choice of the basis of $Harm_{rel,\Bbb Z}^k(M)$, $k=1,2$, since any
different basis is connected by a unimodular transformation leaving the partition function invariant.
\bigskip\bigskip
{\bf 6. Two examples}
\bigskip
In this chapter we want to illustrate the general results obtained in the previous sections with two simple examples.\bigskip

{\bf 6.1. The Maxwell theory on the circle $\Bbb T^1$}
\bigskip

Since any $\Bbb T^1$-bundle over $\Bbb T^1$ is trivial it is possible to choose a global section such that $A_0=0$. Any gauge potential
$A\in\Cal A^{(\Bbb T^1)}$ can be regarded as $\frak t^1$-valued 1-form on the base manifold $\Bbb T^1$. Physically this model does not possess
any dynamics and describes the behavior of time-independent gauge fields on the circle. Furthermore, $\Cal A^{(\Bbb T^1)}$ appears as the
configuration space for the gauge fields in the Hamiltonian formulation of electrodynamics on the manifold $M=\Bbb T^1\times\Bbb R^1$ [37].\par

According to the theorems 4.3 and 4.4 the geometry of the configuration space simplifies as it will be stated in the following:

\proclaim{Corollary 6.1} There exists a principal bundle isomorphism

$$\CD \Cal A^{(\Bbb T^1)} @>\cong >> \Omega ^1(\Bbb T^1;\Bbb R)\otimes \frak
t^1 \\ @VVV @VVV \\
\Cal A^{(\Bbb T^1)}/\frak G_{\ast}^{(\Bbb T^1)} @>\hat{\chi } >> \frak t^1 \\
@VVV @VV \exp V \\
\Cal M_{\ast}^{(\Bbb T^1)} @>(\chi _a)^{-1}>> \Bbb T^1\endCD\tag6.1.1$$ with $\hat{\chi} (\{A\})=\int _{\Bbb T^1} A$, where $\{A\}$ is the
equivalence class with respect to the infinitesimal gauge transformations. Furthermore $(\chi _a)^{-1}([A])=e^{\int _{\Bbb T^1} A}$ according to
(4.1.26). \qed\endproclaim

The 1-dimensional basis of the harmonic differential one forms is provided by the volume form $\rho ^{(1)}=vol_{\Bbb T^1}=-\sqrt{-1}\vartheta$
induced by the canonical flat metric on $\Bbb T^1$. The metric $h$ on $Harm^1(\Bbb T^1;\Bbb R)$ is just $h=2\pi $. Any $\frak t^1$-valued 0-form
$\xi$, respectively any $\frak t^1$-valued 1-form $\tau$ on $\Bbb T^1$ can be rewritten in terms of a Fourier series expansion

$$\xi =\sum\limits _{k\in\Bbb Z}\xi _ke^{2\pi\sqrt{-1}kt},\quad
\tau= \sum\limits _{k\in\Bbb Z}\tau _ke^{2\pi\sqrt{-1}kt}dt.\tag6.1.2$$ Accordingly, all nonzero modes of the gauge field $A$ may always be
gauged away completely by a suitable infinitesimal gauge transformation, whereas its zero mode is affected by topologically non-trivial gauge
transformations, only. \par

The operators $\Delta _0|_{imd_1^{\ast}}$ and $\Delta _1|_{Harm^1(M)^{\perp}}$ possess the same spectrum $\{k^2:k\in\Bbb Z\}$. Let $\zeta
_R(s):=\sum _{k=1}^{\infty}k^{-s}$ be the Riemann $\zeta$-function with the well-known properties $\zeta _R(0)=-\frac{1}{2}$ and
$\frac{d}{ds}|_{s=0}\zeta _R(s)=\ln{(2\pi )^{-1/2}}$, then from (4.2.6) one obtains

$$\zeta (s|\Delta _0|_{imd_1^{\ast}})=2\zeta _R(2s),\tag6.1.3$$
which finally yields

$$\det{(\Delta _0|_{imd_1^{\ast}})}=(2\pi )^2.\tag6.1.4$$
By inserting (6.1.4) into (4.3.4) one obtains for the generating functional
$$\split Z[J]= &(2\pi )^{5/2}\frac{\Theta _1(-J_0|(2\pi )^2\sqrt{-1})}{\Theta _1(0|(2\pi
)^2\sqrt{-1})}\cdot e^{-\frac{1}{4\pi}\sum\Sb k\in\Bbb Z\\
k\neq 0\endSb \frac{|J_k|^2}{k^2}}\\ &\times\frac{\sqrt{-1}}{4\pi J_0(1-J_0^2)}\cdot \left(
e^{-3\pi\sqrt{-1}J_0}+e^{-2\pi\sqrt{-1}J_0}-e^{-\pi\sqrt{-1}J_0}-1 \right) ,\endsplit\tag6.1.5$$ where $J_k$ ($k\in\Bbb Z$) are the Fourier
coefficients of the source $J$ according to the decomposition in (6.1.2). Let us introduce the one-point Green\rq s operator in momentum space
$$\Cal S^{(1)}=\sum\limits _{k\in\Bbb Z}\Cal
S_k^{(1)}e^{2\pi\sqrt{-1}kt}dt,\tag6.1.6$$ which is related to the defining relation (4.3.2) by

$$\Cal S_1(v)=\frac{1}{2\pi}\sum\limits _{k\in\Bbb Z}
\Cal S_k^{(1)}\bar v_k.\tag6.1.7$$ Then (4.3.6) reduces to

$$\Cal S_k^{(1)}=\cases 0, &\text{for $k\neq 0$}\\ 6\pi ^3\sqrt{-1},
&\text{for $k=0$},\endcases\tag6.1.8$$ showing that the one-point Green´s function does not vanish in the topologically nontrivial case.
Analogously, we define the two-point function $\Cal S_{kl}^{(2)}$ in momentum space by

$$\Cal S_2(v^{(1)},v^{(2)})=\frac{1}{2\pi}\sum _{k,l\in\Bbb Z}\Cal
S_{kl}^{(2)}v_{k}^{(1)}\bar v_{l}^{(2)}.\tag6.1.9$$ Then (4.3.7) leads to
$$\Cal S_{kl}^{(2)}=\cases \frac{1}{2\pi k^2}\delta _{kl}, &\text{for $k,l\neq 0$}\\
-2\pi \left( (\frac{17\pi }{12}-\frac{1}{\pi})+ 2\pi\frac{\sum\limits _{m\in\Bbb Z} m^2 e^{-\frac{(2\pi )^2}{2} m^2}}{\sum\limits _{m\in\Bbb Z}
e^{-\frac{(2\pi )^2}{2} m^2}} \right), &\text{for $k=l=0$}.\endcases\tag6.1.10$$ Using (4.1.26) the VEV (4.2.18) of a gauge invariant function
$f$ on $\Cal A^{(\Bbb T^1)}$ simplifies to

$$<f> = \hat{f}_{0}=\int _0^1 dt (\chi _{a}^{\ast}f)(e^{2\pi\sqrt{-1}t}),\tag6.1.12$$ showing
that the zero mode of the gauge field determines the VEV completely. For instance, the
Wilson loop observable $f(A)=e^{\int _{\Bbb T^1}A}$ admits a vanishing VEV, i.e. $<e^{\int _{\Bbb T^1}A}>=0$.\par
\bigskip\bigskip
{\bf 6.2. Abelian gauge theory on two-dimensional manifolds}
\bigskip
On a closed two-dimensional manifold $M$ all principal $\Bbb T^N$-bundles $P$ are uniquely characterized by a $N$-dimensional vector of integers
$m=(m^{\alpha})_{\alpha =1}^N\in\Bbb Z^N$. On each connected component of the space of connections, denoted by $\Cal A^{m}$, let us choose a
background connection $A_0^{m}\in\Cal A^{m}$ such that

$$\int\limits _MF_{A_0^{m}}=2\pi\sqrt{-1}m,\tag6.2.1$$ yields
the Chern number of the corresponding torus bundle $P$. Any two-form on $M$ is proportional to the volume form $vol_M$. In particular,
$F_{A_0^{m}}=\mu vol _M$. Because $d_2^{\ast}F_{A_0^{m}}=0$ it follows that $\mu =\frac{2\pi\sqrt{-1}}{Vol(M)}m$ is a constant . Using (4.2.17)
the partition function for a single topological sector reads

$$Z_{A_0^{m}}^{(m)}=(2\pi )^{b_1N}(\det
h)^{N/2}(\det{\Delta _0|_{imd_1^{\ast}}})^{N}(\det{\Delta _1|_{Harm^1(M)^{\perp}}})^{-N/2}e^{-\frac{1}{2}(2\pi )^2\sum _{\alpha ,\beta
=1}^N\lambda _{\alpha \beta}m^{\alpha}m^{\beta}},\tag6.2.2$$ which sums up to the total partition function on the total configurations space
$\Cal A^{(M)}$

$$Z=(2\pi )^{b_1N}(\det
h)^{N/2}(\det{\Delta _0|_{imd_1^{\ast}}})^{N}(\det{\Delta _1|_{Harm^1(M)^{\perp}}})^{-N/2}\Theta _{N}(0|\frac{2\pi\sqrt{-1}}{Vol(M)}\Lambda
).\tag6.2.3$$ We conclude with some results regarding the VEV of a polynomial of the field strength on the total configuration space $\Cal
A^{(M)}$. The VEV - called $\Cal W_q$ - can be derived from

$$\Cal W_q(\eta _1\ldots ,\eta _q) =\frac{\partial ^q}{\partial
t_1\cdots\partial t_q} \vert _{t_1=\ldots =t_q=0} \ \frac{\sum\limits _{m\in\Bbb Z^N} I^{(m)}(e^{<F_A,\sum _{i=1}^qt_i\eta _i>})}{\sum\limits
_{m\in\Bbb Z^N} Z_{A_0^m}^{(m)}}.\tag6.2.4$$

\proclaim{Corollary 6.2} If the matrix $\lambda$ of couplings between the gauge fields is diagonal, the correlation functions of odd degree
vanish, i.e. $\Cal W_{2k+1}^{(P)}=0$.\qed\endproclaim

\demo{Proof} This result follows directly from the fact that only an odd number of $F_{A_0}$ appears in (6.2.4).\qed\enddemo

For the Maxwell theory ($N=1$) the correlation functions of odd degree henceforth vanish whereas the correlation functions of even degree $q=2k$
yield

$$\split \Cal W_{2k}(&\eta _1\ldots ,\eta _{2k})=\frac{(2\pi )^{2k}}{Vol(M)^{2k}}\frac{\sum\limits
_{m\in\Bbb Z}e^{-\frac{2\pi ^2}{Vol(M)}m^2}m^{2k}}{\Theta _1 (0|\frac{2\pi\sqrt{-1}}{Vol(M)})}\int _M\eta _{\varsigma (1)}\cdots\int _M\eta
_{\varsigma (2k)}\\ &+ \sum _{l=1}^{k-1} \frac{(2\pi)^{2(k-l)}}{2^ll!(2(k-l))!Vol(M)^{2(k-l)}}\frac{\sum\limits _{m\in\Bbb Z}e^{-\frac{2\pi
^2}{Vol(M)}m^2}m^{2(k-l)}}{\Theta _1 (0|\frac{2\pi\sqrt{-1}}{Vol(M)})}\\ &\times\sum _{\varsigma} <\eta _{\varsigma
(1)},\Pi^{Harm^{2}(M)^{\perp}}\eta _{\varsigma (2)}>\cdots <\eta _{\varsigma
(2l-1)},\Pi^{Harm^{2}(M)^{\perp}}\eta _{\varsigma (2l)}>\\
&\times\int _M\eta _{\varsigma (2l+1)}\cdots\int _M\eta _{\varsigma (2k)}\\ &+\frac{1}{2^kk!}\sum _{\varsigma} <\eta _{\varsigma
(1)},\Pi^{Harm^{2}(M)^{\perp}}\eta _{\varsigma (2)}>\cdots <\eta _{\varsigma (2k-1)},\Pi^{Harm^{2}(M)^{\perp}}\eta _{\varsigma
(2k)}>,\endsplit\tag6.2.5$$ where $\Pi^{Harm^{2}(M)^{\perp}}\eta _{\varsigma (j)}=\eta _{\varsigma (j)}-\frac{1}{Vol(M)}(\int _{M}\eta
_{\varsigma (j)})vol_M$ is the projection onto the subspace $imd_2\otimes\frak t^N$. The first term in (6.2.5) is topological as it depends only
on the volume of $M$. (For sake of completeness we refer to [38,39] where the field strength correlation functions in two dimensions have been
calculated by a different method).
\bigskip\bigskip
{\bf 7. Concluding Remarks}\bigskip In this paper we have tried to elaborate on the definition of a measure for gauge theories, which are
affected by the Gribov problem. The starting point has been a modified functional integral, which we have applied to abelian gauge theories
residing on $n$-dimensional compact manifolds $M$ with and without a boundary. In both cases the non-triviality of the bundle of gauge fields
and thus the existence of Gribov ambiguities have been proved. A patching prescription has been developed for the functional integral. We have
used the description of the gauge orbit space as a bundle over a multi-dimensional torus to calculate the partition function, the vacuum
expectation value (VEV) of gauge invariant observables and the Green\rq s functions of the theory. This explicit analysis of the underlying
bundle structure may be also useful for further discussions on abelian gauge theories.
\par
In the particular case of Maxwell theory, our results for both the partition function and the VEV of gauge invariant observables are in exact
agreement with calculations based on the conventional Faddeev-Popov treatment: The volume of the gauge group can be indeed absorbed into a {\it
finite} normalization constant.\par

On the other hand, the Green\rq s functions have been shown to be affected by the non-triviality of the bundle of gauge fields resulting in a
non-vanishing vacuum expectation value of the gauge field. This phenomenon has been related to the first (relative) cohomology of the manifold
$M$.
\bigskip\bigskip
{\bf Acknowledgments}
\bigskip I would like to express my gratitude to H. H\"{u}ffel for his
various comments and his encouragement and to A. Cap for helpful discussions.

\enddocument